\documentclass[twocolumn,nopacs,preprintnumbers]{revtex4}
\usepackage{graphicx}
\usepackage{bm}
\usepackage{color}
\usepackage{xcolor}
\usepackage[normalem]{ulem}
\usepackage{amsfonts}
\usepackage{amsmath}
\usepackage{mathtools}

\usepackage{physics}
\usepackage{subfigure}
\usepackage{pdfpages}

\graphicspath{%
    {converted_graphics/}
    {/}
}

\begin{document}

\title{Quantum Vacuum Sagnac Effect}

\author{Guilherme C. Matos$^1$, Reinaldo de Melo e Souza$^2$, Paulo A. Maia Neto$^1$,  and Fran\c cois Impens$^1$}
\address{$^1$ Instituto de F\'isica, Universidade Federal do Rio de Janeiro, Rio de Janeiro, RJ, 21941-972, Brazil}
\address{$^2$ Instituto de F\'isica, Universidade Federal Fluminense, Niter\'oi, RJ, 24210-346, Brazil}

\begin{abstract}
We report on the quantum electrodynamical analogue of a Sagnac phase induced by the fast rotation of a neutral nanoparticle
onto atomic waves propagating in its vicinity.
The quantum vacuum Sagnac phase is a geometric Berry phase proportional to the angular velocity of rotation. The persistence of a noninertial effect into the inertial frame is also 
analogous to the Aharonov-Bohm effect. Here, a rotation confined to a restricted domain of space gives rise to an atomic phase
even though the interferometer is at rest with respect to an inertial frame. 
By taking advantage of a plasmon resonance, we show that the magnitude of the induced phase can be close to the 
sensitivity limit of state of the art interferometers. The quantum vacuum 
Sagnac atomic phase is a geometric footprint of a dynamical Casimir-like effect.
\end{abstract}

\maketitle

 The rotation of a frame attached to an interferometer with respect to an inertial frame induces an interferometric phase proportional to the angular frequency of the rotation and to the area enclosed by the interferometer. This phenomenon,  known as the Sagnac effect~\cite{Sagnac13}, has several technological applications  such as optical-based Sagnac sensors which have been embarked on aircraft for decades. Three decades ago, it has been extended to matter waves~\cite{Borde89}. Since then, cold-atom gyrometers based on the Sagnac effect~\cite{Gustavson96,Pritchard97} have been constantly improved to outperform their optical equivalents~\cite{Canuel06,RMPCronin09,Sackett20,Ryu2020,Schubert2021} (see Ref.~\cite{Geiger2020} for a recent review).

We propose here to investigate the following closely related scheme  -- what if, instead of rotating the interferometer as a whole, one simply spins a neutral nanoparticle placed between its arms? 
Such question naturally arises as rotation speeds beyond $5\,{\rm GHz}$ are achieved with optically levitated nanoparticles \cite{Ahn2020}.
As in the Casimir effect, we show that  the interaction between the atom and the nanoparticle mediated by the quantum electromagnetic field leads to an atomic phase induced by the particle's spinning. 
We introduce the quantum vacuum Sagnac phase (QVSP) 
as an atom-interferometry footprint of noninertial effects in the quantum vacuum. The QVSP is thus analogous to
the dynamical Casimir effect~\cite{Dalvit2011,Dodonov2020}, but no real photons are emitted in the configuration discussed hereafter.
Indeed, here the relative motions of the interfering wave packets with respect to the sense of the particle's rotation plays a crucial role in 
the atom-particle interaction mediated by the quantum vacuum field.
While dynamical Casimir photons are too scarce to be measurable \cite{MaiaNeto96,Dodonov2021, Dalvit2021}
even when considering a cavity resonance \cite{Lambrecht96,Dodonov96,Schutzhold98,Crocce2001}, 
we show that the magnitude of the QVSP is 
close to the sensitivity limit
of state-of-the-art atom interferometers \cite{Kumiya2016,Geiger2020} when taking into account the record rotation frequencies recently demonstrated with optically levitated nanoparticles~\cite{Reimann2018,Ahn2018,Ahn2020}.

The emission of dynamical Casimir photons out of the quantum vacuum state by a
dielectric sphere undergoing a non-uniform rotation was 
considered in Ref.~\cite{Barton96}. 
The electromagnetic field at finite temperature was predicted to exert a quantum friction
 torque on a neutral dispersive microsphere spinning at a constant rotation frequency~\cite{Manjavacas2010,ManjavacasPRL2010}. 
The QVSP imprinted on the atomic center of mass~(CM) provides an additional insight 
on how the rotation modifies the surrounding quantum electromagnetic field even when no real photons are emitted.

The QVSP is also a consequence of the motion of atomic CM with respect to the nanoparticle.
In the context of atom interferometry, a nonlocal phase associated with pairs of paths (rather than with individual ones) was shown to result from the field-mediated interaction between a moving atom and a material surface~\cite{Nonlocal1,Nonlocal2,Nonlocal3}. Dynamical Casimir emission of photons~\cite{Souza2018,Lo2018,Farias2019,Dolan2020,Agusti2021}, decoherence~\cite{Scheel2012,Souza2016} and 
quantum friction~\cite{Scheel09,Pieplow2013,Reiche2020,Fosco2021,Donaire2016,Farias2020,Lombardo21} also result from the coupling between a moving atom and the quantum electromagnetic field. Given their high sensitivity, atom interferometers are candidates for the first 
experimental demonstration of motional effects in Casimir physics, 
and the QVSP would be particularly appealing for that purpose.

The QVSP is also related to
 the Aharonov-Bohm effect \cite{Aharonov1959}. 
Such connection is well understood in the case of the standard Sagnac effect (see Ref.\cite{ReviewSagnacBerryArxiv21} for a recent review), as 
a rotating referential emulates the presence of magnetic fields thanks to the similarity between Coriolis and Lorentz forces ~\cite{Dalibard11}. This analogy has enabled the production of artificial effective magnetic fields in neutral cold-atom gases set into rotation~\cite{Cornell04}. 
Like the standard Sagnac phase, the QVSP is a geometric phase that can be cast in terms of an effective magnetic field.
In addition, the QVSP can be seen as the Aharonov-Bohm-like counterpart of the Sagnac effect. Indeed, in the Aharonov-Bohm experiment, a magnetic field confined to a solenoid imprints a phase in a region free of magnetic fields. Here, we show that a rotation confined to a domain of space imprints
a phase on matter waves probing quantum vacuum fluctuations 
outside the rotating region.

\begin{figure}
\includegraphics[scale=0.25]{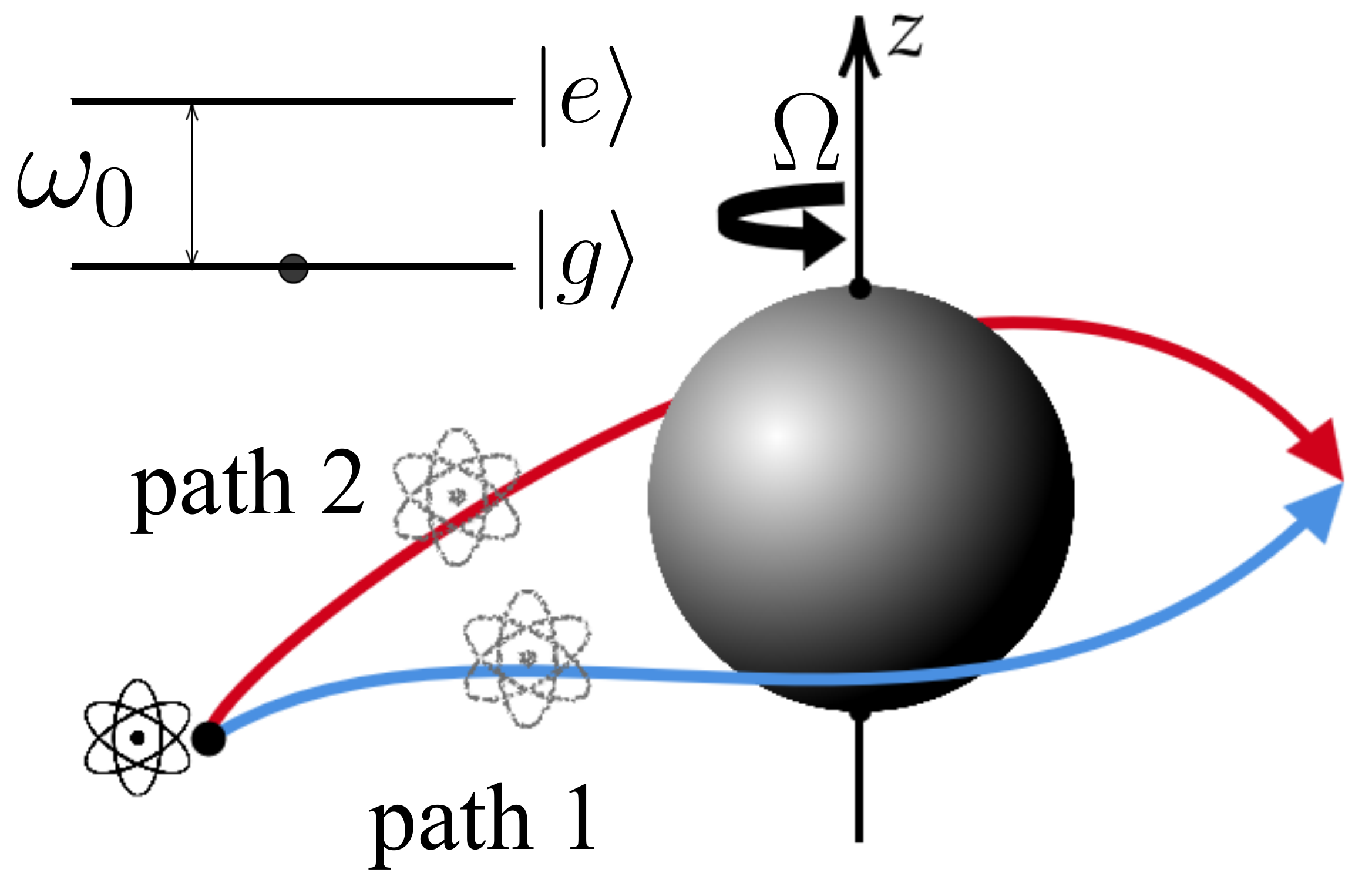}
\caption{Scheme of the quantum vacuum Sagnac interferometer. The CM of a ground-state atom propagates as a quantum superposition of two wave packets around a spinning neutral nanoparticle (angular frequency $\Omega$).
 \label{scheme}
}
\end{figure}

For simplicity, we consider a nanoparticle rotating around an axis of symmetry with constant angular velocity $\Omega.$ In this case, 
the modification of the surrounding quantum field arises from the frequency dependence of the particle dielectric constant.
We consider a 2-level atom in the ground state 
interacting with the quantum vacuum field.
The atom CM is in a quantum superposition of two wave packets
that propagate in the vicinity of the spinning nanoparticle as indicated in Fig.~\ref{scheme}. We show that the resulting 
QVSP is geometric, i.e. independent of the atomic velocity~{\cite{Miniatura92,Gauguet13}} in the limiting case of very narrow wave packets. Furthermore, we express the QVSP as the circulation of a geometric vector field, analog to the vector potential in the Aharonov-Bohm effect,  along the interferometer paths. The effect can be enhanced by considering nanoparticles with a plasmon resonance~\cite{PlasmonMetallic2012} in order to optimize the material dispersion at the atomic transition frequency. 

\textit{Motional van der Waals (vdW) atomic phase.}  
 We consider a moving atom interacting with the rotating nanoparticle between the initial/final times
$t=\mp T/2$. The atomic waves acquire a phase associated with the dipolar interaction $\hat{H}_{\rm dip}= -\hat{\mathbf{d}} \cdot \hat{\mathbf{E}},$ with 
$\hat{\mathbf{d}}$ representing the atomic dipole moment operator. The
 electric field operator  $\hat{\mathbf{E}}$ is
  taken at the instantaneous average atomic position $\mathbf{r}_k(t)=\langle\hat{ \mathbf{r}}(t) \rangle_k$
  for each wave packet $k.$ 
We evaluate the phase difference $\Delta \phi_{12}$ accumulated by 
the coherent superposition state of
two narrow atomic wave packets following the two distinct paths $\mathcal{P}_1 =[\mathbf{r}_1(t)],$  $\mathcal{P}_2 =[\mathbf{r}_2(t)].$ 
Up to second order in perturbation theory, this phase difference reads~\cite{Nonlocal3}   
  \begin{eqnarray}
\label{totalphase}
\Delta \phi_{12} & = & \varphi_{11}-\varphi_{22}+\varphi_{12}-\varphi_{21} \\
\label{eq:linearresponse}
\mathcal{\varphi}_{kl} & = & \frac {1} {4}\iint_{-\frac{T}{2}}^{\frac{T}{2}} dt\,  dt'  \left[ g^{H}_{\hat{\mathbf{d}}}(t,t')  \mathcal{G}^{R,S}_{\hat{\mathbf{E}}}(\mathbf{r}_k(t),t;\mathbf{r}_l(t'),t') \right. \nonumber \\
& \: & \left. \qquad \qquad+ g^{R}_{\hat{\mathbf{d}}}(t,t')  \mathcal{G}^{H,S}_{\hat{\mathbf{E}}}(\mathbf{r}_k(t),t;\mathbf{r}_l(t'),t') \right] 
\end{eqnarray}
The contributions $\mathcal{\varphi}_{kl}$ for $k=l$ and $k \neq l$ correspond  to local and non-local phases, respectively. 
In the concrete applications discussed later on, the local phases $\phi_k \equiv \varphi_{kk}$ will play a more important role. We have used the trace of the retarded Green's function for the scattered electric field 
$\mathcal{G}^{R,S}_{\hat{\mathbf{E}}}(\mathbf{r},t;\mathbf{r}',t')= {\rm Tr} \left[ \boldsymbol{G}^{R,S}_{\hat{\mathbf{E}}}(\mathbf{r},t;\mathbf{r}',t') \right]$, which captures how electrodynamical propagation is modified by the presence of the nanoparticle (scatterer) placed at the origin.  
Likewise, the trace $\mathcal{G}^{H,S}_{\hat{\mathbf{E}}}$  of the Hadamard Green's function represents the change in the field fluctuations 
induced by the presence of the nanoparticle. 
The retarded Green's function of a vectorial operator $\hat{\mathbf{O}}(t)$ is defined as the averaged commutator $\boldsymbol{G}_{\hat{\mathbf{O}} \: ij}^R (t,t') = \frac {i} {\hbar} \Theta(t-t') \langle [ \hat{O}_i(t),\hat{O}_j(t')] \rangle$ with $\Theta(\tau)$ denoting the Heaviside function. 
The Hadamard Green's function corresponds to the average value of the anti-commutator  $\boldsymbol{G}_{\hat{\mathbf{O}} \: ij} ^H (t,t')  = \frac {1} {\hbar} \langle \{ \hat{O}_i(t), \hat{O}_j(t') \} \rangle.$ 

The first term on the r.-h. s. of (\ref{eq:linearresponse}) 
accounts for the electric field response to dipole fluctuations, while the second one corresponds to the dipole response to vacuum
fluctuations modified by the presence of the nanoparticle. 
The dipole Hadamard Green's function is isotropic 
and has the analytical form 
$g^{H}_{\hat{\mathbf{d}}\: ij}(t,t')=\alpha_0^{\rm A} \omega_0 \cos \omega_0(t-t') \delta_{ij}$
for a 2-level model. Here, 
 $\alpha_0^{\rm A}$ represents 
the static polarizability and  $\omega_0$
is the transition frequency.  
We focus on the nonretarded vdW regime, for which the atom-particle distance  $r(t)$ is much smaller than the transition wavelength $\lambda_0=2 \pi c / \omega_0 $. As shown below, the QVSP 
is maximized in the immediate vicinity of the rotating nanoparticle, which turns the vdW regime more interesting 
for experimental implementations.

We now consider the retarded Green's function for the scattered electric field  $ \boldsymbol{G}^{R, (S)}_{\hat{\mathbf{E}} \: ij}(\mathbf{r},t;\mathbf{r}',t') 
.$ This function corresponds to the $i$-th component of the electric field at position $\mathbf{r}$ and time $t$ induced by an instantaneous point dipole oriented along the $j$-th direction at position $\mathbf{r}'$ and time $t'$ after scattering at the nanoparticle at some intermediate time $t''$ such that $t'<t''<t$. From now on, 
we assume that the nanoparticle is very small and neglect multipolar contributions beyond the electric dipolar one. The retarded Green's function in the frequency domain
can then be expressed in terms of the electric polarizability tensor $\boldsymbol{\alpha^{\Omega}}(\omega)$ of the rotating nanoparticle
as
\begin{equation}
\boldsymbol{G}_{\hat{\mathbf{E}}}^{R, S}(\mathbf{r},\mathbf{r}',\omega) = \boldsymbol{G}^0(\mathbf{r},\mathbf{0},\omega) \cdot
 \boldsymbol{\alpha^{\Omega}}(\omega)\cdot \boldsymbol{G}^0(\mathbf{0},\mathbf{r}',\omega)
 \label{eq:generalGreenFunction}
\end{equation}
The free-space 
 retarded Green's function
 for the electric field~\cite{HeitlerBook} becomes frequency-independent in the nonretarded vdW regime $\boldsymbol{G}^{0}_{ij}(\mathbf{r},\mathbf{r}',\omega)\approx (3  R_i R_j / R^{2} -  \delta_{ij})/(4\pi\epsilon_0R^{3})$ with $\mathbf{R}=\mathbf{r}-\mathbf{r}'$. 
In the absence of rotation, any direction orthogonal to the symmetry axis of the nanoparticle is a principle axis of the polarizability tensor $\boldsymbol{\alpha}^0(\omega)$ with an eigenvalue denoted by $\tilde\alpha(\omega)$. Rotation around the symmetry axis leads to a nondiagonal correction $\delta\boldsymbol{\alpha^{\Omega}}(\omega)_{lm}=\boldsymbol{\alpha^{\Omega}}_{lm}(\omega)-\boldsymbol{\alpha}^0_{lm}(\omega)\approx i \tilde{\alpha}'(\omega)  \sum_n  \epsilon_{lmn}  \Omega_n$~\cite{Manjavacas2010} which lies at the heart of the QVSP. We have assumed a nonrelativistic rotation with angular velocity $\boldsymbol{\Omega}$, $\tilde{\alpha}'(\omega)$ is the frequency derivative of the polarizability eigenvalue, and $\epsilon_{lmn}$ denotes the Levi-Civitta tensor components.

\textit{QVSP induced by a rotating nanoparticle.}  We provide first a heuristic discussion of the QVSP derived below. This phase arises from the scattering of virtual photons on the spinning nanoparticle, which carry back to the atom a trace of the particle rotation. 
The QVSP
captures the noninertial footprint on the quantum electromagnetic field as the atom probes fluctuations in the vicinity of the particle. 
In this sense, it constitutes a \textit{dynamical Casimir-like} effect. However, as a nanoparticle spinning at constant velocity produces no radiation~\cite{Barton96,Manjavacas2010}, the QVSP does not rely on the presence of real dynamical Casimir photons -- nor does it rests on open-quantum system dynamics~\cite{Lombardo06} responsible for quantum friction~\cite{Farias2020,Lombardo21}.

The problem under consideration involves very different time scales, listed below from the slowest to the fastest in the vdW regime:  the
time-of-flight $T$
of the atomic CM in the vicinity of the rotating particle and the period of rotation (a fraction of nanosecond), the inverse of the atomic transition frequency $2 \pi/\omega_0$, the response time of the rotating particle due to dispersion, and finally the light travel time between the moving atom and the particle, $r/c$. 
Such hierarchy of time scales allows us to take several approximations. The dominant contribution in Eq.~\eqref{eq:linearresponse} comes from intervals $t-t'$ of the order of the response time of the nanoparticle, enabling us to neglect the CM acceleration and take $\mathbf{r}_k(t') \simeq \mathbf{r}_k(t)- (t-t') \mathbf{v}_k(t)$ in Eqs.~(\ref{eq:linearresponse}) and \eqref{eq:generalGreenFunction}.

 In addition, the displacement of the atomic CM during $t-t'$ is
  much smaller than the wavelengths of field modes contributing to the electric field Green's functions.
  Thus, we Taylor expand the latter around the  position $\mathbf{r}_k(t)$. 
  We start by deriving the local contribution to the QVSP and 
  define $\phi^{\rm \Omega}_{k }$ as the $\Omega-$dependent contribution to the phase $\phi_{k}$ in Eq.~\eqref{eq:linearresponse}.
We write it as the sum $\phi^{\rm \Omega}_{k } =  \phi^{\rm \Omega}_{{\rm qs}, k}+\phi^{\rm \Omega}_{{\rm mot},k}$ of a quasistatic and of a motional contribution. 
  The former is obtained by taking identical arguments for the retarded and advanced positions ($\mathbf{r}=\mathbf{r}'=\mathbf{r}_k(t)$) in the electric 
  Green's function, while the latter involves instead the gradient of the Green's functions and the instantaneous atomic CM velocity. 
  
  The quasistatic contribution $\phi^{\rm \Omega}_{{\rm qs},k}$
   vanishes by symmetry considerations~\cite{Supplementary}. 
   The QVSP thus arises exclusively from the atomic motion during the electrodynamical delay time $t-t'$ associated with the exchange of virtual photons between the atom and the nanoparticle.
We write  $\phi^{\rm \Omega}_{ k} = \phi^{\rm \Omega}_{{\rm mot}, k}= \phi^{\rm \Omega, dip}_{k}+ \phi^{\rm\Omega, f}_{k}$
   as the sum of contributions from dipole and field fluctuations, respectively.
    They correspond to the first and second terms in
 Eq.~\eqref{eq:linearresponse}.
We  use the condition $\omega_0 T\gg 1$ to derive
\begin{equation}\label{eq:dipoleSagnacphasefrequency}
 \phi^{\rm \Omega, dip}_{k} =    -\frac {\omega_0 \alpha_0^{\rm A}} {4}  
 \int_{-T/2}^{T/2} \! \! \! \! \! \! dt   \: \mathbf{v}_k(t) \cdot    \partial_{\omega} \nabla_{\mathbf{r}'}  \mbox{Im} \left[  \delta\mathcal{G}^{R,S}_{\hat{\mathbf{E}}}(\mathbf{r},\mathbf{r}';\omega)  \right] 
\end{equation}
where the spatial and frequency derivatives are taken at $\mathbf{r}'=\mathbf{r}=\mathbf{r}_k(t)$ and $\omega=\omega_0,$ respectively. 
Here, $\delta\mathcal{G}^{R,S}_{\hat{\mathbf{E}}}$ represents the $\boldsymbol{\Omega}-$dependent contribution to the scattered Green function, which is obtained 
 by taking  $\delta\boldsymbol{\alpha^{\Omega}}(\omega)$ instead of the 
 full tensor $\boldsymbol{\alpha}(\omega)$  in  Eq.~(\ref{eq:generalGreenFunction}). 
The frequency derivative captures the time delay associated with the nanoparticle response to the field (in the form of an induced dipole) during the atomic motion.
On the other hand, the delay associated with the light propagation time from the atom to the nanoparticle is 
negligible within the vdW approximation. Accordingly, we take the nonretarded vdW approximation for the free-space Green's function  $\boldsymbol{G}^0(\mathbf{r},\mathbf{r}',\omega)$ 
when deriving the scattered field propagator $\delta \mathcal{G}^{R,S}_{\hat{\mathbf{E}}}(\mathbf{r},\mathbf{r}';\omega) $ from Eq.~\eqref{eq:generalGreenFunction}.

The contribution from field fluctuations $\phi^{\rm \Omega, f}_{k}$ is given by an expression similar to~\eqref{eq:dipoleSagnacphasefrequency} in terms of the retarded Green's function for the dipole (polarizability) and the Hadamard Green's function for the field. 
Using the fluctuation-dissipation theorem  $  \nabla_{\mathbf{r}'}  \mathcal{G}_{\hat{\mathbf{E}}}^{H,S}(\mathbf{r},\mathbf{r}';\omega) = 2\, {\rm sgn}(\omega)\, \mbox{Im} [ \nabla_{\mathbf{r}'}  \mathcal{G}_{\hat{\mathbf{E}}}^{R,S}(\mathbf{r},\mathbf{r}';\omega) ],$ 
with ${\rm sgn}(\omega)$ denoting the sign function, 
we find a contribution to the local QVSP identical to the dipolar one.
We write the final result for the local QVSP as a geometric integral~\cite{Supplementary}, which is the main result of this Letter:
\begin{eqnarray}
\label{eq:quantumSagnac1}
\phi^{\rm \Omega}_{k}  =  \frac92\,
 \frac { \omega_0 \alpha_0^{\rm A}  \tilde{\alpha}_R''(\omega_0)} { (4 \pi \epsilon_0)^2} \int_{\mathcal{P}_k}  d \mathbf{r}  \:    \cdot   \: \frac { \mathbf{\Omega} \times \mathbf{r}} {r^8},
 \end{eqnarray}
where  $\tilde{\alpha}_R(\omega_0)$ is the real part of the nanoparticle's polarizability eigenvalue.
The integral is performed along the interferometer path $\mathcal{P}_k$ delimited by the initial/final positions $\mathbf{r}_k(\mp T/2).$  
As in the standard Sagnac effect~\cite{Dalibard11,Cornell04}, the QVSP given by Eq.~\eqref{eq:quantumSagnac1} is a 
geometric phase that can be cast as the line integral of an effective vector potential proportional to the angular velocity $\Omega.$ The QVSP~\eqref{eq:quantumSagnac1} possess all the distinctive features of a geometric phase: it is independent of the velocity magnitude, but changes sign when the direction of propagation is reversed.

As an important insight in its geometric nature, one can show~\cite{Supplementary} that the QVSP is indeed a Berry phase~\cite{Berry1,Berry2} of the full quantum system ``2-level atom+field'' undergoing a unitary and adiabatic quantum evolution steered by the atomic position. Because of the dipole interaction, the instantaneous ground state of this quantum system changes continuously as the atom propagates nearby the spinning nanoparticle, following a quantum trajectory which depends on the interferometer path. By integrating the corresponding Berry connection, one retrieves exactly the local QVSP~(5).

 The correction associated to eletrodynamical retardation is given by~\cite{Supplementary} $\phi^{\rm \Omega}_{(c), k}  =   3
  \omega_0 \alpha_0^{\rm A}  \tilde{\alpha}_R'(\omega_0) / [(4 \pi \epsilon_0)^2 c^2] \int_{\mathcal{P}_k}  d \mathbf{r}  \:    \cdot   \: \frac { \mathbf{\Omega} \times \mathbf{r}} {r^6}$ and is negligible for the example discussed below.
   In the case of a finite temperature $\theta,$ the local QVSP is multiplied by the factor 
   $\coth(\hbar\omega_0/2k_B\theta)$ ($k_B=$ Boltzmann constant), which is very close to one  
for any realistic example of atomic transition. In other words, the contribution from 
thermal fluctuations is negligible and the QVSP is a genuine quantum vacuum effect. 
     
 \textit{Examples of interferometer designs.}  To illustrate our findings, we calculate the phase for two specific geometries:
   either a circle of radius $R$ centered on the nanoparticle, or two parallel straight lines enclosing the particle.
   In both cases we assume that ${\bf \Omega}=\Omega\,\mathbf{\hat z}$ is orthogonal to the plane containing the trajectories.
    The first arrangement is a  textbook example of a Sagnac interferometer, whereas the latter corresponds to a more
  realistic situation in atom interferometry~\cite{Cronin04,Vigue09,Vigue11}. 
   One finds $\phi^{\rm \Omega}_{\{ r=R \}} = 9 \pi \ell^6_{\Omega}/ R^6$ for the circular trajectory when the senses of rotation of atom and nanoparticle coincide; and $\phi^{\rm \Omega}_{\{ x=x_1,  - L/2 \leq y \leq + L/2, z=z_1 \}} \approx 45 \pi \ell^6_{\Omega}\, x_1/ (32 r_{\perp}^7) $ with $r_{\perp}=(x_1^2+z_1^2)^{1/2}$ for a straight segment of length $L$
satisfying the condition  $|x_1|, |z_1| \ll L/2 \ll \lambda_0$ for consistency with the vdW approximation.
We have introduced the characteristic length scale $\ell_{\Omega}= [\omega_0 \alpha_0 \alpha_R''(\omega_0)  \Omega / (4 \pi \epsilon_0)^2]^{1/6}.$

\textit{Non-local contributions to the QVSP difference.} 
 We now discuss the non-local QVSP contribution $ \varphi^{\Omega}_{12}-\varphi^{\Omega}_{21} $ corresponding to the $\Omega$-dependent part of the non-local phases in Eq.~\eqref{totalphase}. In a different context, cross-talks of the interferometer paths can influence significantly the motional phases~\cite{Nonlocal1,Nonlocal2,Nonlocal3} or decoherence rates~\cite{Mazzitelli03,Scheel2012} induced by the quantum vacuum.  While the nonlocal QVSP
difference vanishes for circular trajectories, it is nonzero when considering 
 two parallel straight line trajectories with the nanoparticle at the midpoint ($x_2=-x_1$  and $z_1=z_2=0$).
In this case, the non-local QVSP contribution  has an opposite sign with respect to the local one,  thus reducing the total QVSP difference to $\Delta \phi^{\Omega}_{12}=63 \pi \ell^6_{\Omega}\,{\rm sgn}(x_1)/ (32 x_1^6).$
  For this geometry, the non-local contribution represents a sizable part of the total QVSP difference.

\textit{Estimation of the QVSP for finite-width atomic wave packets and a spherical nanoparticle.} 
We estimate the magnitude of the  QVSP in a practical interferometer implementation.  Specifically, we consider a Mach-Zehnder configuration where path $1$ flies near a rotating nanosphere of radius $a$, and path $2$ evolves far away from the nanosphere. A similar interferometer geometry was used in~\cite{Cronin04,Vigue09,Vigue11} for the measurement of quasistatic vdW phases. In this setup, the QVSP receives only a local contribution from path $1$, namely $\Delta \phi_{12}^{\Omega}=\phi_1^{\Omega}$. In order to enhance the  QVSP, we investigate materials for which the polarizability exhibits a sharp frequency dependence at the atomic transition frequency. 
 
 For this purpose, we use a plasmon resonance~\cite{PlasmonMetallic2012} and seek metals for which
 the polarizability $\tilde{\alpha}(\omega)=(4 \pi \epsilon_0) a^3 [\epsilon(\omega)-1]/[\epsilon(\omega)+2]$ is maximized by reducing 
  $\epsilon(\omega) +2$ at the  atomic transition frequency $\omega_0.$
 Within the Drude model, the dielectric constant reads $\epsilon(\omega)= 1 - \omega_P^{2}/[ \omega (\omega + i \gamma)],$ where  $\omega_P$ 
 is the plasma frequency and $\gamma$ is the inverse of the electronic relaxation time. 
 The plasma resonance in the dipole approximation is at $\omega_{\rm res} = \omega_P / \sqrt{3}.$ Since we want to maximize $\tilde{\alpha}_R''(\omega)$ rather than  $\tilde{\alpha}'(\omega),$
  we need the atomic transition frequency to be slightly shifted with respect to  $\omega_{\rm res}.$ 
  We consider Na atoms,  for which the
  static polarizability is $\alpha_0^{\rm A}/(4 \pi \epsilon_0) = 2.4\times  10^{-29} {\rm m^{3}}$ 
and the dominant transition  ($3s_{1/2} - 3p_{3/2}$) has a frequency  $\omega_ 0 = 3.198 \times 10^{15} \:{\rm rad/s}$  \cite{Cronin10}.
  We take the plasma frequency of the nanosphere to be 
 $\omega_{\rm P}=5.549 \times 10^{15} {\rm rad/s}$ so as to maximize $\tilde{\alpha}_R''(\omega_ 0).$
 Such fine tuning can be achieved, for instance, from the size dependence of the plasmon resonance in nanospheres \cite{Derkachova2016,Karimi2019}.
 Our value for  $\omega_{\rm P}$ is also very close, within less than $1\%,$ to the bulk plasma frequency of potassium \cite{Blaber09}. Accordingly, we take 
 the relaxation frequency of potassium $\gamma=2.795\times 10^{13}\,{\rm rad/s}$ in our numerical estimation~(Fig.~2). 
  
 \begin{figure}[htpb]
\includegraphics[width=8cm]{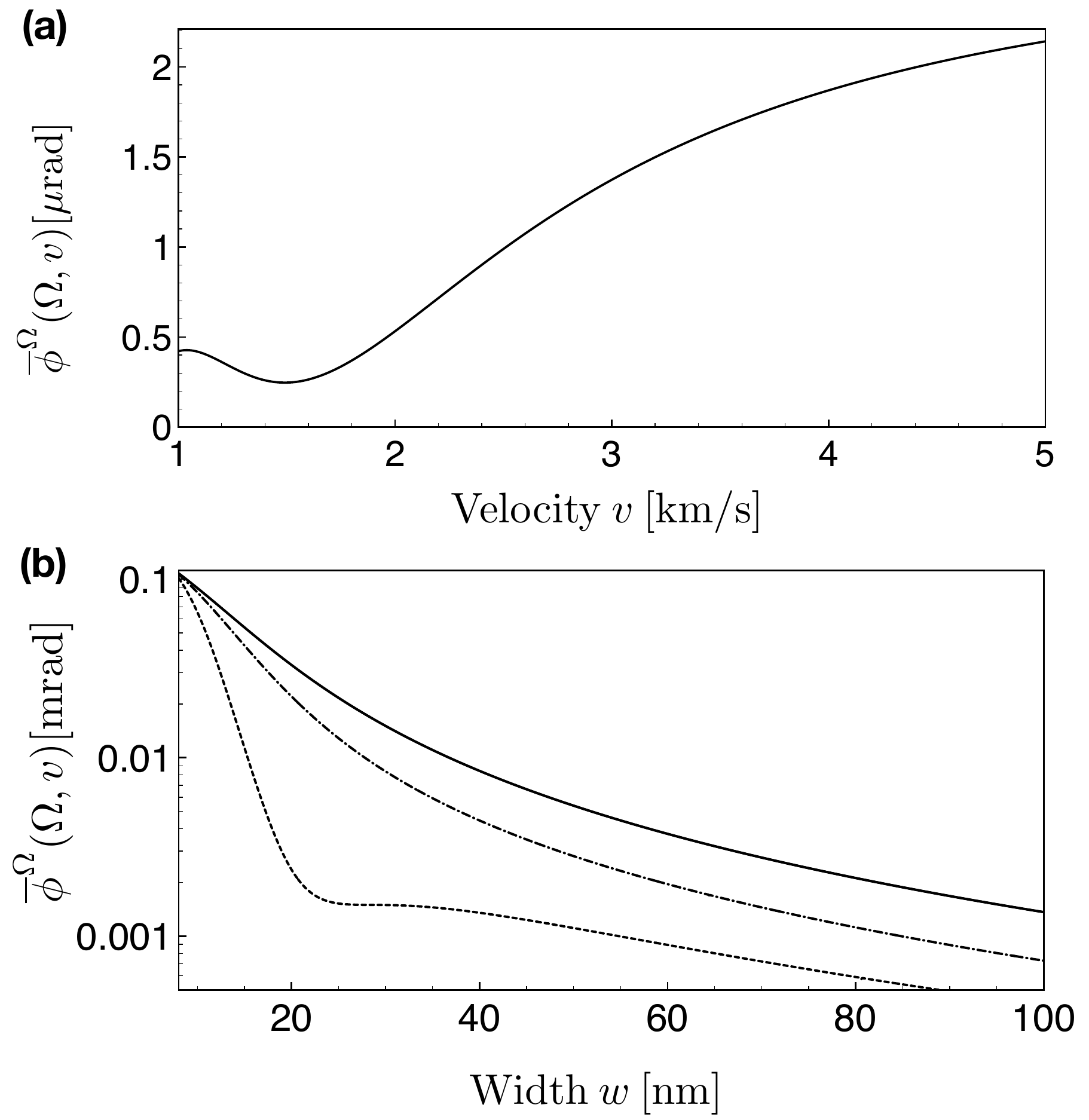}
   \caption{Average QVSP versus (a) velocity and (b)  width  of the Gaussian atomic beam used in the interferometer.
    Parameters: (a)  $a= 50 \: {\rm nm}$, $w = 100 \: {\rm nm}$. (b)  $a=  35 \: {\rm nm}$,  $v = 3 \: {\rm km/s}$ (dotted line),  $v = 4 \: {\rm km/s}$ (dash-dotted line) and $v = 5 \: {\rm km/s}$ (solid line). In (a,b) we have considered a 2-level Na atom with the static polarizability  $\alpha_{0}^{\rm A}= (4 \pi \epsilon_0) \times 2.4 \times 10^{-29}{\rm m}^3$ and transition frequency $\omega_{0} = 3.198 \times 10^{15} \:{\rm rad/s}$. We have taken a nano-sphere of Potassium spinning at the angular velocity $\Omega= 2\pi \times 5 \: {\rm GHz}$. }
	\label{fig:2}
\end{figure}

We average the atomic phase
over the  transverse wave packet widths following the procedure of Refs.~\cite{Cronin04,Vigue09,Vigue11} for (quasistatic) vdW phases. 
The phase acquired by atoms flying along the trajectory  $\{ x'=x, - L/2 \leq y' \leq L/2 ,z'=z \}$ at a constant velocity $\mathbf{v}=v \: \hat{\mathbf{y}},$ is the sum of the vdW phase $\phi^{\rm vdW}(x,z,v),$
 as derived by integration of the instantaneous vdW potential along the trajectory,
 and the QVSP  $\phi^{\Omega}(x,z)$ : $\phi(\Omega,x,z,v)=\phi^{\rm vdW}(x,z,v)+ \phi^{\Omega}(x,z).$ The 
 known result \cite{Cronin04} for the vdW phase
follows from Eqs.~(\ref{eq:linearresponse}) and (\ref{eq:generalGreenFunction}) in the quasistatic limit:
$\phi^{\rm vdW}(x,z,v) \simeq  9 \pi \alpha_0 \omega_0 \tilde{\alpha}_R(\omega_0)/[4.(4 \pi \epsilon_0)^2 v r_{\perp}^5]$  with $r_{\perp}=(x^2+z^2)^{1/2}$. There are two important differences between the vdW phase and the QVSP. The former is dynamical and thus inversely proportional to the velocity, while the latter is geometric and thus velocity-independent. On the other hand, the vdW phase is unaffected by the nanosphere's spinning, while the  QVSP is proportional to the  angular velocity $\Omega.$ 
 
 The experimentally accessible phase reads  $\overline{\phi}(\Omega,v)={\rm Arctan} \left[ \overline{\sin( \phi(\Omega,x,z,v) )} / \overline{\cos( \phi(\Omega,x,z,v) )}  \right],$ where the averaging is performed over the transverse coordinates $x,z$. We consider
  a Gaussian atomic wave packet of transverse width $w$ and longitudinal velocity $v$. In practice, one may gradually increase the nanosphere rotation  in order to 
   isolate
   the average QVSP as the $\Omega-$dependent part of the total average phase:
 $\overline{\phi}^{\Omega}(\Omega,v)\equiv\overline{\phi}(\Omega,v)-\overline{\phi}(0,v).$  
  In spite of the geometric nature of   $\phi^{\Omega}(x,z)$, the presence of the dynamical phase $\phi^{\rm vdW}(x,z,v)$ together with the averaging procedure
 involving trigonometric functions turn $\overline{\phi}^{\Omega}(\Omega,v)$ velocity-dependent in the case of finite-width wave packets.
 
   We take the angular velocity $\Omega= 2\pi \times 5 \: {\rm GHz}$ recently achieved with 
    optically-levitated nanoparticles~\cite{Ahn2020}.  We also consider thin atomic beams of width $w \leq 100 \, {\rm nm}$ centered at the edge of the spinning particle. Such collimation may be obtained by using diffraction through a nano-grating~\cite{Vigue11} placed in the vicinity of the spinning particle, or by tight focusing techniques considered for atom lithography~\cite{Prentiss92}. Prospective focusing techniques~\cite{Martin21a,Martin21b} show indeed that atomic beam widths $w \simeq 8\, {\rm nm}$ may be attained. We plot  $\overline{\phi}^{\Omega}(\Omega,v)$
    versus velocity in Fig.~\ref{fig:2}a for a nanosphere of radius $a=50\,{\rm nm}$ and an atomic wave packet of width $w=100\,{\rm nm}.$
    Fig.~\ref{fig:2}b  presents the variation of  $\overline{\phi}^{\Omega}(\Omega,v)$ with the wave packet width, for $a=35\,{\rm nm}$
    and for different velocities of a few km/s.
    The averaged QVSP tends to increase with the 
 atomic velocity and is enhanced by
  atomic beam focusing. However, 
  $\overline{\phi}^{\Omega}(\Omega,v)$  may be significantly attenuated around specific velocities or width values under the influence of the quasistatic vdW phase. To avoid the detrimental effect of the vdW potential,
  we consider fast atomic beams, with velocities comparable to those of Ref.~\cite{Vigue11}.
   Fig.~\ref{fig:2}b shows that an  average QVSP $\overline{\phi}^{\Omega}(\Omega,v) \simeq 0.1\, {\rm mrad}$ is attained
for $w = 8  \: {\rm nm}$. Such value is close to the current phase sensitivity limit in atom interferometry~\cite{Geiger2020,Salvi2018}.

\textit{Conclusions.}   We have shown that the fast rotation of a nanoparticle imprints a geometric phase analogous to a Sagnac phase on a ground-state
 atom propagating in its vicinity. 
  The persistence of a noninertial effect beyond the region  where the rotation actually occurs is reminiscent of the Aharonov-Bohm effect, with the vector 
  potential yielding a finite atomic phase in a region free of magnetic field. We have assumed the interferometer to be at rest with respect to an inertial frame.
  Thus, the noninertial effects are exclusively mediated by the quantum vacuum field  through the scattering on a rotating nanoparticle. The resulting
  QVSP is a dynamical Casimir-like modification of the atomic phase, whose observation might be more at hand than the detection of dynamical Casimir photons. Quantum dipolar and field fluctuations contribute equally to the QVSP, which
  can be enhanced by using materials exhibiting a plasmon resonance near the atomic transition frequency.
  The QVSP might become within reach of experimental observation given the state of the art in atom interferometry and 
  nanorotors.

 \begin{acknowledgments}
F.I. thanks Jean Dalibard for stimulating initial discussions, and A. Z. Khoury, M. B. da Silva Neto for discussions on geometric phases. F.I. and P.A.M.N. are grateful to Ryan O. Behunin and Claudio Ccappa Ttira for a previous collaboration on non-local atomic phases. 
 We acknowledge funding from the Brazilian agencies 
 Conselho Nacional de Desenvolvimento Cient\'{\i}fico e Tecnol\'ogico (CNPq) (409994/2018-9),
Coordena\c c\~ao de Aperfei\c coamento de Pessoal de N\'{\i}vel Superior
 (CAPES), and Funda\c c\~ao de Amparo \`a Pesquisa do Estado do Rio de Janeiro (FAPERJ) (202.874/2017, 210.242/2018 and 210.296/2019).
 P.A.M.N. also thanks partial finantial support from 
 Instituto Nacional de Ci\^encia e Tecnologia de Fluidos Complexos (INCT-FCx) and 
Funda\c c\~ao de Amparo \`a Pesquisa do Estado de S\~ao Paulo  (FAPESP) (2014/50983-3). 

\end{acknowledgments}

\newpage

\begin{widetext}
 
\section*{SUPPLEMENTAL MATERIAL}


In this Supplemental Material, we provide additional details on the derivation of the Quantum Vacuum Sagnac Phase (QVSP) arising from the interaction between a 2-level atom in the ground state and a spinning nano-particle. We provide an alternative derivation of the local QVSP based on the Berry connection of the full quantum system ``2-level atom+field'' steered by the atomic position. We show the consistency of our model with the well-known van-der-Waals potential in the quasistatic regime. We also obtain the non-local contributions to the quantum vacuum Sagnac phase, and detail the averaging procedure used for finite atomic wave-packets.\\


 We remind here for convenience a few notations of the main text. The scattered electric-field Green's function $\boldsymbol{G}_{\hat{\mathbf{E}}}^{R, S}(\mathbf{r},\mathbf{r}',\tau) = \frac {1} {2 \pi} \int d \omega \boldsymbol{G}_{\hat{\mathbf{E}}}^{R, S}(\mathbf{r},\mathbf{r}',\omega) e^{-i \omega \tau} $
is obtained from the following tensor product
\begin{equation}
\boldsymbol{G}_{\hat{\mathbf{E}}}^{R, S}(\mathbf{r},\mathbf{r}',\omega) = \boldsymbol{G}^0(\mathbf{r},\mathbf{0},\omega) .
 \boldsymbol{\alpha}^{\Omega}(\omega) . \boldsymbol{G}^0(\mathbf{0},\mathbf{r}',\omega)
 \label{eq:generalGreenFunction}
\end{equation}
The free-field Green's function is given by
\begin{equation}
\label{eq:freepropagatorFourier}
\boldsymbol{G}^0(\mathbf{r},\mathbf{r}',\omega)= \frac {e^{i k R}}  {4 \pi \epsilon_0 R^3}  \boldsymbol{T}(\mathbf{R},\omega),
\end{equation}
 with $T_{ij}(\mathbf{R},\omega)=(k^2 R^2 + i k R -1) \delta_{ij} - (k^2 R^2 + 3 i k R -3)R^{-2}  R_i R_j$ and 
 $k=\omega/c$,$\mathbf{R}=\mathbf{r}-\mathbf{r}'.$ We shall often use the van-der-Waals~(vdW) approximation of the free-field Green's function
\begin{equation}
\label{eq:vdWpropagator}
\boldsymbol{G}^{\rm vdW}(\mathbf{r},\mathbf{r}') = \frac {1} {4 \pi \epsilon_0 R^{3}} \boldsymbol{T}^{0} (\mathbf{R}) \qquad {\rm with} \qquad
T^{0}_{ij} (\mathbf{R})= 3  \hat{R}_i \hat{R}_j -  \delta_{ij}
\end{equation}
 and the unit vector $\hat{\mathbf{R}}=\mathbf{R}/R$. 
    
 The free-field Green's function $\boldsymbol{G}^0(\mathbf{r},\mathbf{r}',\tau)$ obtained from~\eqref{eq:freepropagatorFourier} is a retarded propagator, as well as the nano-particle polarizability response $ \mathbf{\alpha}^{\Omega}(\tau)$. By Eq.~\eqref{eq:generalGreenFunction}, the scattered Green's function $\boldsymbol{G}_{\hat{\mathbf{E}}}^{R, S}(\mathbf{r},\mathbf{r}',\tau)$ is a convolution of these functions, and is thus also a retarded propagator.
 
 In practice, we use only the trace of scattered electric-field Green's functions $\mathcal{G}_{\hat{\mathbf{E}}}^{R (H), S}(\mathbf{r},\mathbf{r}',\tau)= {\rm Tr} \left[ \boldsymbol{G}_{\hat{\mathbf{E}}}^{R (H), S}(\mathbf{r},\mathbf{r}',\tau)  \right].$ We assume from now on thermal equilibrium and the zero-temperature limit. The Fluctuation-Dissipation Theorem~(FDT) yields:
\begin{eqnarray}
 \mathcal{G}^{H,S}_{\hat{\mathbf{E}}}(\mathbf{r},\mathbf{r};\omega) \! & \! = \! & \! 2 {\rm sgn}(\omega) \mbox{Im} \left[ \mathcal{G}^{R,S}_{\hat{\mathbf{E}}}(\mathbf{r},\mathbf{r};\omega)  \right] 
 \label{eq:FDT}
\end{eqnarray}
${\rm sgn}(\omega)$ is a sign function, i.e. ${\rm sgn}(\omega)=1$ (-1) for $\omega >0$ ($\omega<0$).  A similar equality holds for the dipole Green's functions.

\section{Obtention of the dispersive phase in the quasistatic approximation}

We first use our model to obtain the dispersive phase associated to the interaction between a moving atom and a nano-particle (spinning or at rest). We take here the quasistatic approximation, namely we neglect the effects of the atomic motion, one retrieves the expected vdW phase. Beyond this consistency check, this derivation is used in Section~\ref{sec:averageQVSP} to estimate the average Quantum Vacuum Sagnac Phase~(QVSP) for finite atomic wave-packets.

  We start by writing the local phases  of Eq.(1) in the main text [noted here $\phi_{{\rm  qs}}$ to emphasize the quasistatic approximation]:
\begin{eqnarray}
\label{phi_disp} 
\phi_{{\rm  qs},k}=  \! \frac {1} {4} \int_{-T/2}^{T/2} dt  \int_{-T/2}^{T/2} dt'   \left[ \frac {} {} g_{\hat{d}}^H(t,t') \mathcal{G}_{\hat{\mathbf{E}}}^{R,S}(\mathbf{r}_k(t),\mathbf{r}_k(t);t-t')
 +    g_{\hat{d}}^R(t,t')  \:   \mathcal{G}_{\hat{\mathbf{E}}}^{H,S}(\mathbf{r}_k(t),\mathbf{r}_k(t);t-t') \frac {} {}  \right] 
\end{eqnarray}
where we have taken the quasistatic approximation by setting $\mathbf{r}_k(t') \simeq \mathbf{r}_k(t)$. 
By Parseval's theorem, in the long-time limit $\omega_0 T \gg1$ this phase reads:
\begin{equation}
\label{phi_dispFourier}
\phi_{{\rm  qs},k} =  \! \frac {1} {4  ( 2 \pi)} \int_{-T/2}^{T/2} \! \! \! \! \! \! dt  \int_{- \infty}^{+ \infty} \! \! \!  \! \! \!  d \omega   \left[ \frac {} {} g_{\hat{d}}^H(-\omega) \mathcal{G}_{\hat{\mathbf{E}}}^{R,S}(\mathbf{r}_k(t),\mathbf{r}_k(t);\omega) +    g_{\hat{d}}^R( -\omega )  \:   \mathcal{G}_{\hat{\mathbf{E}}}^{H,S}(\mathbf{r}_k(t),\mathbf{r}_k(t);\omega) \frac {} {}  \right] 
\end{equation}
The first and second term on the r.h.s. correspond respectively to the quasistatic contributions of dipole and field fluctuations. To simplify notations, we omit in the rest of this Section the path label and take simply the phase along $\mathcal{P} \equiv [\mathbf{r}(t)]$.


\subsection{Contribution of dipole fluctuations to the dispersive phase}

 Using the expression of the dipole Hadamard Green's function of the dipole in frequency
\begin{equation}
g^{H}_{\hat{d}}(\omega)=\pi \alpha^{\rm A}_0 \omega_0 [\delta(\omega-\omega_0)+\delta(\omega+\omega_0)] \, ,
\label{eq:HadamardDipole}
\end{equation}
one can write the phase $\phi^{\rm dip}_{{\rm qs}}$ related to dipole fluctuations as
\begin{equation}
\label{eq:phase_vdW_dipole1}
\phi^{\rm dip}_{{\rm qs}}  =  \! \frac {\alpha^{\rm A}_0 \omega_0} {4} \int_{-T/2}^{T/2} \! \! \! \! \! \! dt {\rm Re} \left[ \mathcal{G}_{\hat{\mathbf{E}}}^{R,S}(\mathbf{r}(t),\mathbf{r}(t);\omega_0) \right]
\end{equation}
For the considered van-der-Waals regime, we substitute in Eq.~\eqref{eq:generalGreenFunction} the free-field propagator $\mathbf{G}^0(\mathbf{r},\mathbf{r}',\omega)$ by its unretarded approximation~\eqref{eq:vdWpropagator}:
\begin{equation}
\phi^{\rm dip}_{{\rm qs}}    =  \! \frac {\alpha_0^{\rm A} \omega_0}  {4 (4 \pi \epsilon_0)^2} \int_{-T/2}^{T/2} \! \! \! \! \! \! dt  \frac {1} {r(t)^6} \sum_{l,m,n} \left(  \delta_{lm} - 3 \frac {r_l(t) r_m(t)} {r^{2}(t)}  \right) \boldsymbol{\alpha}_{R \: mn}(\omega_0) \left(  \delta_{nl} - 3 \frac {r_n(t) r_l(t)} {r^{2}(t)}  \right) 
\end{equation}
where the summation is done over the three space coordinates $\{x,y,z\}$). The static polarizability $\boldsymbol{\alpha}^{0}(\omega)$ is given in terms of its dispersive and dissipative part as $ \boldsymbol{\alpha}(\omega) =  \boldsymbol{\alpha}_R(\omega) + i  \boldsymbol{\alpha}_I(\omega).$ For a non-rotating object, the polarizability tensor is diagonal, so that $\boldsymbol{\alpha}_{R,  mn}(\omega_0) = \tilde{\alpha}_R(\omega_0) \delta_{mn}.$ The summation can be done as
\begin{eqnarray}
\label{eq:identity6}
\sum_{l,m,n}  \left(  \delta_{lmn} - 3 \frac {r_l(t) r_m(t)} {r^{2}(t)}  \right) \delta_{mm} \left(  \delta_{nl} - 3 \frac {r_n(t) r_l(t)} {r^{2}(t)}  \right)  & = & \left(  3- 2 \times 3  +3 \times 3 \right) = 6
\end{eqnarray}
Finally, the quasistatic phase is given by the time integral as the atom moves along the path $\mathcal{P}=[\mathbf{r}(t)]$
\begin{equation}
\phi^{\rm dip}_{{\rm qs}}     =  \! \frac {3  \omega_0  \alpha^{\rm A}_0  \tilde{\alpha}_R(\omega_0)} {2  (4 \pi \epsilon_0)^2} \int_{-T/2}^{T/2}  \frac {dt} {r(t)^6} 
\end{equation}

\subsection{Dispersive phase associated to the electric field fluctuations} 
 
The phase contribution arising from the field fluctuations reads, by virtue of the FDT~\eqref{eq:FDT} 
\begin{equation}
\phi^{\rm f}_{{\rm qs}}  =  \! \frac {1} {2  ( 2 \pi)} \int_{-T/2}^{T/2} \! \! \! \! \! \! dt  \int_{- \infty}^{+ \infty} \! \! \!  \! \! \!  d \omega \: {\rm sgn}(\omega) g_{\hat{d}}^R( -\omega )  \:   \mbox{Im} \left[ \mathcal{G}^{R,S}_{\hat{\mathbf{E}}}(\mathbf{r}(t),\mathbf{r}(t);\omega)  \right] 
\label{eq:vdWfieldfluctuation}
\end{equation} 
The retarded Green's functions $\mathcal{G}^{R,S}_{\hat{\mathbf{E}}}(\mathbf{r},\mathbf{r};\omega),g^{R}_{\hat{\mathbf{d}}}(\omega)$ have real/imaginary parts which are even/odd functions of the frequency $\omega$.  By parity, only the real part ${\rm Re} [g_{\hat{d}}^R( \omega )] = \frac 1 2 (g_{\hat{d}}^R( \omega )+g_{\hat{d}}^R( -\omega ))  $ contributes to the integral~\eqref{eq:vdWfieldfluctuation}. The retarded dipole Green's function reads $g^{R}_{\hat{d}}(\tau)= \alpha_0^{A} \omega_0 \Theta(\tau)\sin \omega_0\tau$. Using the relation
  \begin{equation}
 \int_0^{+ \infty} d\tau e^{i (\omega-\omega_0) \tau} =  i \: \mathcal{P} \left( \frac {1} {\omega-\omega_0} \right) + \pi \delta (\omega-\omega_0)
 \end{equation}
 where $\mathcal{P}$ is the principal value, we obtain
\begin{equation}
\label{eq:dipoleRetardedGreenFunction} 
g^{R}_{\hat{d}}(\omega)= \frac 1 2 \alpha_0^{\rm A} \omega_0 \left[  \mathcal{P} \left( \frac {1} {\omega+\omega_0} \right) - \mathcal{P} \left( \frac {1} {\omega-\omega_0} \right) \right] + \frac {i \pi} {2}  \alpha_0^{\rm A} \omega_0 [ \delta(\omega-\omega_0) - \delta(\omega+\omega_0) ) ].
\end{equation}
 As a consistency check, one has $g^{R}_{\hat{d}}(-\omega)=g^{R}_{\hat{d}}(\omega)^*$. 
Taking the vdW approximation in the scattered field propagator, namely
\begin{eqnarray}
 {\rm Im} \left[ \mathcal{G}^{R,S}_{\hat{\mathbf{E}}}(\mathbf{r},\mathbf{r}';\omega)  \right] =  {\rm Tr} \left[  \boldsymbol{G}^{\rm vdW}(\mathbf{r},\mathbf{0}) .
\boldsymbol{\alpha}_{I}(\omega) . \boldsymbol{G}^{\rm vdW}(\mathbf{0},\mathbf{r}')  \right]
\end{eqnarray}
and using the isotropy of the polarizability tensor $\boldsymbol{\alpha}_{I, ij} (\omega)= \tilde{\alpha}_{I}(\omega) \delta_{ij},$ we arrive at the following expression for the quasistatic phase
\begin{equation}
\phi^{\rm f}_{{\rm qs}}   =  \! \frac {3 \alpha_0^{\rm A} \omega_0} {4 \pi (4 \pi \epsilon_0)^2}  \left[ \int_{-\infty}^{+\infty}  d \omega \: {\rm sgn}(\omega)  \tilde{\alpha}_I(\omega)   \mathcal{P} \left( \frac {1} {\omega+\omega_0} - \frac {1} {\omega-\omega_0} 
 \right) \right]  \int_{-T/2}^{T/2} \frac {dt} {r(t)^6}   
\end{equation}
where we have used again the identity~\eqref{eq:identity6}. The frequency integral can be simplified by parity arguments. Introducing
\begin{equation}
\label{eq:effectiveimaginarypolar}
\overline{\alpha_{I}}(\omega_0) = \frac {1} {\pi} \mathcal{P} \left[ \int_{-\infty}^{+\infty} d \omega  \: {\rm sgn}(\omega)  \frac { \tilde{\alpha}_I(\omega) } {\omega +\omega_0}  \right] 
\end{equation}
Summing these contributions, we obtain the total quasistatic dispersive phase
\begin{equation}
\label{eq:genericvdWphase}
\phi_{{\rm  qs}} =  \! \frac {3 \omega_0 \alpha_0^{\rm A}  (\tilde{\alpha}_{R}(\omega_0)+\overline{\alpha_{I}}(\omega_0 ))} {2 (4 \pi \epsilon_0)^2}  \int_{-T/2}^{T/2} \frac {dt} {r(t)^6}   
\end{equation}
In the numerical examples considered below $\overline{\alpha_{I}}(\omega_0 ) \approx \tilde{\alpha}_R(\omega_0)$, so that field and dipole fluctuations contribute almost equally to the quasistatic phase.


\subsection{Consistency with the vdW potential}

Finally, we retrieve the vdW atom-atom interaction by replacing the scatterer polarizability with the retarded atom Green's function~\eqref{eq:dipoleRetardedGreenFunction}. Precisely, in this Section only, one sets $\tilde{\alpha}(\omega)=g_{\hat{d}}^{R}(\omega)$. As $\mathcal{P} \int dx \frac {\delta(x+x_0)} {x+x_0}=  \lim_{\epsilon \rightarrow 0^{+}} \left[ \int_{-\infty}^{-x_0-\epsilon} dx+\int_{-x_0+\epsilon}^{+ \infty} dx \right] \left( \frac {\delta(x+x_0)} {x+x_0} \right)=0,$ one finds $\overline{\alpha}_{I}(\omega_0) =  \frac 1 4  \alpha_0^A$. Similarly, $\tilde{\alpha}_R(\omega_0)= \int d\omega \delta(\omega-\omega_0) {\rm Re}[ g_{\hat{d}}^{R}(\omega)] =  \frac 1 4 \alpha_0^A .$ Thus, the quasistatic dispersive phase reads in this case
\begin{equation}
\phi_{{\rm  qs}, k}^{\rm at.-at.} =  \! \frac {3 \omega_0 \alpha_0^{\rm A 2} } {4 (4 \pi \epsilon_0)^2}  \int_{-T/2}^{T/2} \frac {dt} {r(t)^6}   
\end{equation}
This is consistent with the expected vdW phase, given by the temporal integration $\phi_{{\rm  qs}, k}^{\rm at.-at.} = - \frac 1 \hbar \int_{-T/2}^{T/2} V_{\rm vdW}(\mathbf{r}(t))$ of the vdW potential between two ground-state atoms~\cite{ThiruBook} $V_{\rm vdW}(\mathbf{r})= - 3 \hbar \omega_0 \alpha_0^{\rm A 2}  / [4  (4 \pi \epsilon_0)^2 r^6].$
 
 \section{Obtention of the Quantum Vacuum Sagnac Phase}
 
  The local QVSP $\phi^{\rm \Omega}_k$  corresponds to a local phase [$\varphi_{kk}$, Eq.(1) of the main text]:
 \begin{eqnarray}
\label{phi_general}
\phi^{\rm \Omega}_{k}  =  \! \frac {1} {4 } \int_{-T/2}^{T/2} dt  \int_{-T/2}^{T/2} dt'   \left[ \frac {} {} g_{\hat{d}}^H(t,t') \delta \mathcal{G}_{\hat{\mathbf{E}}}^{ R,S}(\mathbf{r}_k(t),\mathbf{r}_k(t');t-t')
 +    g_{\hat{d}}^R(t,t')  \:  \delta \mathcal{G}_{\hat{\mathbf{E}}}^{H,S}(\mathbf{r}_k(t),\mathbf{r}_k(t');t-t') \frac {} {}  \right] 
\end{eqnarray}
in which one retains only the scattered electric-field Green's functions $ \delta \mathcal{G}_{\hat{\mathbf{E}}}^{ R(H),S}(\mathbf{r}_k(t),\mathbf{r}_k(t');t-t')$ associated to the $\Omega$-dependent polarizability tensor
 \begin{equation}
 \delta \alpha^{\Omega}_{lm}(\omega) = i \sum_n \epsilon_{lmn} \Omega_n      \tilde{\alpha}' (\omega)
 \label{eq:nondiagonalpolarizability}
 \end{equation}
$\epsilon_{klm}$ is the Levi-Civitta tensor. The $\Omega$-dependent retarded electric-field Green's function reads
\begin{equation}
\label{eq:spinningscatteredGreen}
\delta \boldsymbol{G}_{\hat{\mathbf{E}}}^{R, S}(\mathbf{r},\mathbf{r}',\omega) =  \boldsymbol{G}^0(\mathbf{r},\mathbf{0},\omega) .
\delta \boldsymbol{\alpha}^{\Omega}(\omega) . \boldsymbol{G}^0(\mathbf{0},\mathbf{r}',\omega)
\end{equation}
and the corresponding Hadamard Green's function $\delta \boldsymbol{G}_{\hat{\mathbf{E}}}^{H, S}(\mathbf{r},\mathbf{r}',\omega) $ is obtained from the FDT~\eqref{eq:FDT} in the low-temperature limit.

In order to account for motional effects, we write the CM atomic position as: $\mathbf{r}_k(t') \simeq \mathbf{r}_k(t)- (t-t') \mathbf{v}_k(t)$ and use a perturbative expansion of the Green's tensor. The QVSP then admits the formal expression $\phi^{\rm \Omega} =  \phi^{\rm \Omega}_{\rm qs}+\phi^{\rm \Omega}_{\rm mot}$ where  $\phi^{\rm \Omega}_{\rm qs}$ and $\phi^{\rm \Omega}_{\rm mot}$ represent respectively a quasistatic and a motional contribution.The latter reads in the long-time limit $\omega_0 T \gg1$
\begin{eqnarray}
\phi^{\rm \Omega}_{\rm mot, k}  =  \! \frac {-1} {4} \int_{-T/2}^{T/2} \! \! \! \! dt \:   \mathbf{v}_k(t) \cdot   \int_{- \infty}^{+ \infty} \!   \! \! \!  d \tau   \left[ \frac {} {} \tau g_{\hat{d}}^H(\tau)  \left. \nabla_{\mathbf{r}'}  \delta \mathcal{G}_{\hat{\mathbf{E}}}^{R,S}(\mathbf{r},\mathbf{r}';\tau) \right|_{\mathbf{r}=\mathbf{r}'=\mathbf{r}_k(t)}
 +  \tau  g_{\hat{d}}^R( \tau )  \left.   \nabla_{\mathbf{r}'} \delta  \mathcal{G}_{\hat{\mathbf{E}}}^{H,S}(\mathbf{r},\mathbf{r}';\tau) \right|_{\mathbf{r}=\mathbf{r}'=\mathbf{r}_k(t)} \frac {} {}  \right] 
 \label{eq:dynamicalSagnac}
\end{eqnarray}

 \subsection{Cancellation of the quasistatic contribution to the QVSP}

The quasistatic dipole contribution to the Sagnac phase is given by an expression analog to Eq.~\eqref{eq:phase_vdW_dipole1}, up the replacement of the scattered electric field propagator by its $\Omega$-dependent contribution $\delta \mathcal{G}_{\hat{\mathbf{E}}}^{R, S}(\mathbf{r}(t),\mathbf{r}(t),\omega)$. Nevertheless, by using Eqs.~(\ref{eq:nondiagonalpolarizability},\ref{eq:spinningscatteredGreen}), one sees that 
$\delta \mathcal{G}_{\hat{\mathbf{E}}}^{R, S}(\mathbf{r}(t),\mathbf{r}(t),\omega) =   0$
 by contraction of the antisymmetric Levi-Civitta tensor $\epsilon_{klm}$ with the symmetric tensor $ G^0_{jk}(\mathbf{r}(t),\mathbf{0},\omega) G^0_{lj}(\mathbf{0},\mathbf{r}(t),\omega).$ As a result, the Sagnac phase has no quasistatic dipole contribution.
The quasistatic field contribution to the Sagnac phase, analog to Eq.~\eqref{eq:vdWfieldfluctuation}, vanishes by the same argument. Thus, there is no quasistatic contribution to the QVSP, and $ \phi^{\rm \Omega}_{k} = \phi^{\rm \Omega}_{\rm mot, k}.$

 \subsection{General expressions for the QVSP associated to dipole and field fluctuations} 

\subsubsection{Contribution of dipole fluctuations to the QVSP}

From Eq.~\eqref{eq:dynamicalSagnac}, one obtains the quantum Sagnac phase associated to dipole fluctuations:
 \begin{eqnarray}
\phi^{\rm \Omega, dip}_{k}  =  
 \frac {-1} {4} \int_{-T/2}^{T/2} \! \! \! \! dt \:   \mathbf{v}_k(t) \cdot   \int_{- \infty}^{+ \infty} \!   \! \! \!  d \tau    \tau g_{\hat{d}}^H(\tau)  \left. \nabla_{\mathbf{r}'} \delta \mathcal{G}_{\hat{\mathbf{E}}}^{R,S}(\mathbf{r},\mathbf{r}';\tau) \right|_{\mathbf{r}=\mathbf{r}'=\mathbf{r}_k(t)}
\end{eqnarray}
 Using $\int d\tau \tau e^{-i(\omega+\omega') \tau}= (2 \pi) i \delta'(\omega+\omega'),$ one can recast this phase contribution as a frequency integral
 \begin{eqnarray}
\phi^{\rm \Omega, dip}_{k}  =  
 \frac {i} {4 (2 \pi)} \int_{-T/2}^{T/2} \! \! \! \! dt \:   \mathbf{v}_k(t) \cdot   \int_{- \infty}^{+ \infty} \!   \! \! \!  d \omega  g_{\hat{d}}^H(-\omega) \left. \nabla_{\mathbf{r}'} \frac {d} {d \omega} \delta \mathcal{G}_{\hat{\mathbf{E}}}^{R,S}(\mathbf{r},\mathbf{r}';\omega)  \right|_{\mathbf{r}=\mathbf{r}'=\mathbf{r}_k(t)}
\end{eqnarray} 
We use again parity considerations, precisely that the real/imaginary part of the electric field Green's function are even/odd functions of the frequency $\omega$. The Hadamard dipole Green's function is of even parity. Thus, only the imaginary part of the electric field Green's function contributes, and by using Eq.~\eqref{eq:HadamardDipole} one obtains:
 \begin{eqnarray}
\phi^{\rm \Omega, dip}_{k} =  
 \frac {- \omega_0 \alpha_0^{\rm A}} {4} \int_{-T/2}^{T/2} \! \! \! \! dt \:   \mathbf{v}_k(t) \cdot  {\rm Im} \left[   \left.  \nabla_{\mathbf{r}'}   \frac {d} {d \omega} \delta \mathcal{G}_{\hat{\mathbf{E}}}^{R,S}(\mathbf{r},\mathbf{r}';\omega) \right] \right|_{\mathbf{r}=\mathbf{r}'=\mathbf{r}_k(t),\omega=\omega_0}
\label{eq:Sagnacdipole}
\end{eqnarray}  

\subsubsection{Contribution of field fluctuations to the QVSP}

The corresponding phase contribution
    \begin{equation}
\phi^{\rm \Omega,f }_{k}  =  \! \frac {i} {4 ( 2 \pi)} \int_{-T/2}^{T/2} \! \! \! \! \! \! dt   \: \mathbf{v}_k(t) \cdot  \int_{- \infty}^{+ \infty} \! \! \!  \! \! \!  d \omega \:  \: g_{\hat{d}}^R( -\omega )  \:    \left. \nabla_{\mathbf{r}'}  \frac {d} {d \omega} \delta \mathcal{G}^{H,S}_{\hat{\mathbf{E}}}(\mathbf{r},\mathbf{r}';\omega)  \right|_{\mathbf{r}=\mathbf{r}'=\mathbf{r}_k(t)}
\end{equation}
can be written, by virtue of the FDT~\eqref{eq:FDT}, as: 
     \begin{equation}
\phi^{\rm \Omega,f }_{k}  =  \! \frac {i} {2 ( 2 \pi)} \int_{-T/2}^{T/2} \! \! \! \! \! \! dt   \: \mathbf{v}_k(t) \cdot  \int_{- \infty}^{+ \infty} \! \! \!  \! \! \!  d \omega \: \:    {\rm sgn}(\omega) \: g_{\hat{d}}^R( -\omega )  \:  \left.  \nabla_{\mathbf{r}'}  \frac {d} {d \omega} \mbox{Im} \left[ \delta \mathcal{G}^{R,S}_{\hat{\mathbf{E}}}(\mathbf{r},\mathbf{r}';\omega)  \right]  \right|_{\mathbf{r}=\mathbf{r}'=\mathbf{r}_k(t)}
\end{equation}
The dipole and field retarded Green's functions have real/imaginary parts which are respectively of even/odd parity in the frequency. Taking into account the sign function and the frequency derivative, one sees that only the imaginary part of the retarded dipole Green's function contributes. Using Eq.~ \eqref{eq:dipoleRetardedGreenFunction}, one obtains an expression of  $\phi^{\rm \Omega,f }_{k}$ identical to~\eqref{eq:Sagnacdipole}, i.e. $\phi^{\rm \Omega,f }_{k} =\phi^{\rm \Omega, dip}_{k}$. Thus the total QVSP reads
   \begin{equation}
   \label{eq:QVSPgeneral1}
\phi^{\rm \Omega}_{k}  =  \! \frac {-    \omega_0 \alpha_0^{\rm A}} {2} \int_{-T/2}^{T/2} \! \! \! \! \! \! dt   \: \mathbf{v}_k(t) \cdot   \: \:     {\rm Im} \left[   \left.  \nabla_{\mathbf{r}'}   \frac {d} {d \omega} \delta \mathcal{G}_{\hat{\mathbf{E}}}^{R,S}(\mathbf{r},\mathbf{r}';\omega) \right] \right|_{\mathbf{r}=\mathbf{r}'=\mathbf{r}_k(t),\omega=\omega_0}
\end{equation}

 \subsection{Expression of the QVSP as a geometric integral} 

The frequency derivative of the $\Omega$-dependent scattered Green's function yields three separate terms in the QVSP~\eqref{eq:QVSPgeneral1}, associated respectively to the delay of the material response and to the light travel time between the moving atom and the nano-particle. To leading-order in the small parameter $\omega_0 r/c$, one has:
\begin{eqnarray}
  \left. \frac {d} {d \omega} {\rm Im} \left[ \delta \mathcal{G}_{\hat{\mathbf{E}}}^{R,S}(\mathbf{r},\mathbf{r}';\omega) \right] \right|_{\omega=\omega_0}  =  \sum_{j,l,m,n} \epsilon_{lmn} \Omega_n & &  \left[ \frac {} {} 
G_{jl}^{\rm vdW}(\mathbf{r},\mathbf{0}) \tilde{\alpha}_R''(\omega_0) G_{mj}^{\rm vdW}(\mathbf{0},\mathbf{r'})  \label{eq:GreensFunctionDerivative} \right. \\
&  \: &  \left.  +   G'^{0}_{jl}(\mathbf{r},\mathbf{0},\omega_0) \tilde{\alpha}_R'(\omega_0) G^{\rm vdW}_{mj}(\mathbf{0},\mathbf{r}') 
+  G_{jl}^{\rm vdW}(\mathbf{r},\mathbf{0}) \tilde{\alpha}_R'(\omega_0)   G'^{0}_{mj}(\mathbf{0},\mathbf{r}',\omega_0) \frac {} {} \right]   \nonumber 
\end{eqnarray} 
The first line term yields the leading QVSP contribution associated to the material response time [noted $\phi^{\rm \Omega}_{{\rm vdW},k}$], while the second line contributions capture corrections to the QVSP arising from retardation effects [noted $\phi^{\rm \Omega, dip}_{(c),k}$].

From Eqs.~(\ref{eq:QVSPgeneral1},\ref{eq:GreensFunctionDerivative}), one finds the non-retarded QVSP contribution arising from the nano-particle response time: 
 \begin{eqnarray}
\phi^{\rm \Omega}_{{\rm vdW},k}  =  
 \frac {\omega_0 \alpha_0^{\rm A}   \tilde{\alpha}_R''(\omega_0)} {4 (4 \pi \epsilon_0)^2} \sum_{j,l,m,n} \epsilon_{lmn} \Omega_n \int_{\mathcal{P}_k}  \frac {d \mathbf{r}} {r^6} \:    \cdot    T^0_{jl}(\mathbf{r})   \nabla_{\mathbf{r}}  T^0_{mj}(\mathbf{r}) 
 \end{eqnarray}   
 We have used the vdW propagator~\eqref{eq:vdWpropagator} and $d \mathbf{r} = \mathbf{v}_k(t) dt $ in Eq.~\eqref{eq:Sagnacdipole} in order to obtain a geometric integral along the path $\mathcal{P}_k$. We also use $\frac {\partial} {\partial r_i}  T^0_{lj}(\mathbf{r}) = 3 ( \delta_{il} \hat{r}_j + \delta_{ij} \hat{r}_l  - 2 \hat{r}_i \hat{r}_j \hat{r}_l)/r.$ Contributions to the differential   $d \mathbf{r}  \cdot \sum_j T_{jl}^0(\mathbf{r})   \nabla_{\mathbf{r}}  T^0_{mj}(\mathbf{r}) $ symmetric with respect to the indices $l,m$ actually disappear by contraction with the Levi-Civitta tensor $\epsilon_{lmn}.$  Thus, the unretarded QVSP is given by the integral along the path $\mathcal{P}_k$ 
    \begin{eqnarray}
    \label{eq:QSVP}
\phi^{\rm \Omega}_{{\rm vdW},k}  =  
 \frac { 9  \omega_0 \alpha_0^{\rm A}     \tilde{\alpha}_R''(\omega_0)} {2 (4 \pi \epsilon_0)^2} \int_{\mathcal{P}_k}  d \mathbf{r}  \:    \cdot   \: \frac { \mathbf{\Omega} \times \mathbf{r}} {r^8} 
 \end{eqnarray}

We now proceed to estimate the contribution to the QVSP $\phi^{\rm \Omega, dip}_{(c),k}$ associated to retardation. We first write the derivative of the free propagator, keeping only the leading-order in the small parameter $k R \ll 1$:
\begin{equation}
G'^{0}_{\rm \: ij}(\mathbf{r},\mathbf{r}',\omega) \simeq \frac  {\omega} {c^2 R}   \left( \hat{R}_i \hat{R}_j  + \delta_{ij}\right) 
\end{equation}

 From Eqs.~(\ref{eq:Sagnacdipole},\ref{eq:GreensFunctionDerivative}), one obtains the contribution to the QVSP arising from retardation effects
  \begin{equation}
\phi^{\rm \Omega}_{(c),k} =  \frac {3 \omega_0^2 \alpha_0^{\rm A}    \tilde{\alpha}_R'(\omega_0)} {(4 \pi \epsilon_0)^2 c^2} \int_{\mathcal{P}_k}  d \mathbf{r}  \:    \cdot   \: \frac { \mathbf{\Omega} \times \mathbf{r}} {r^6} 
\end{equation} 
In the example discussed below retardation effect represent a very small correction, so that in practice $\phi^{\rm \Omega}_{k} \simeq \phi^{\rm \Omega}_{{\rm vdW},k}.$

   \section{Expression of the quantum vacuum Sagnac phase as a Berry phase.}

To establish the connection between the motional phase and the Berry phase, we consider the full quantum system ``2-level atomic sytem+EM field'' undergoing a unitary evolution in presence of the spinning nano-particle and use a perturbative treatment of the dipolar interaction.

For the considered two-level quantum emitter propagating in the quantum vacuum, one notes $\hat{H}_0= \hat{H}_A+\hat{H}_F$ the interaction-free Hamiltonian, with $\hat{H}_A,\hat{H}_F$ are the atomic and the free-field Hamiltonians respectively.The dipole interaction potential reads $\hat{V}(\mathbf{r})= - \hat{\mathbf{d}} \cdot \hat{\mathbf{E}}(\mathbf{r})$ in the Schr\"odinger picture, so that the full Hamiltonian is given by $\hat{H}= \hat{H}_0+\hat{V}(\mathbf{r})$. Switching to the interaction picture, the dipolar interaction becomes time-dependent
\begin{equation}
\label{eq:interaction_picture}
\hat{V}(\mathbf{r},t)= e^{ \frac i \hbar \hat{H}_0 t } \hat{V}(\mathbf{r}) e^{- \frac i \hbar \hat{H}_0 t }.
\end{equation}
The system is assumed to be initially in the ground atomic state atom and with a field in the vacuum state, so that the global quantum state is noted  $| 0 \rangle = | g \rangle_A  \otimes | 0 \rangle_F.$ 
This initial state is coupled through the dipolar interaction to the one-photon states $| n \rangle =  | e \rangle_A \otimes | 1_{n} \rangle_F$ associated to each possible mode $n$ of the EM field in presence of the boundary condition (the spinning nano-particle). These are eigenstates of the free Hamiltonian $\hat{H}_0$, i.e.  $\hat{H}_0 | n \rangle = \hbar \omega_n |n \rangle.$  The index $n$ represents indeed a continuous parameter corresponding typically to a wave-vector and a polarization. We use the short-hand notation $\sum_n$ for a sum over all possible one-photon quantum states - this sum corresponds actually to a continuous integral. It is convenient to use these notations, as we do not need to evaluate the sum in order to establish the desired correspondence between the QVSP and a Berry phase.

\subsection{Expression of the QVSP in terms of transition amplitudes to one-photon states}

We first write the dispersive phase $\phi_k$ along a path $k$ (including motional effects) as a second-order term in the Dyson series~\cite{Nonlocal3} 
\begin{equation}
\label{eq:dispersiveDyson}
\phi_k = {\rm Re} \left[  \frac {{i}} {\hbar^2} \int_{-T/2}^{T/2} dt \int_{-T/2}^t dt'  \langle 0 | \hat{V}(\mathbf{r}_k(t),t) \hat{V}(\mathbf{r}_k(t'),t') | 0 \rangle \right]
\end{equation}
with the dipolar potential $\hat{V}(\mathbf{r},t)$ in the interaction picture~(\ref{eq:interaction_picture}). Eq.(\ref{eq:dispersiveDyson}) is indeed equivalent to the starting point of the QVSP derivation in the main text [Eq.(1) of the Letter], by virtue of the following relation
  ${\rm Re} \left[ \frac {i} {\hbar^2}  \langle 0 | (-\hat{d}(t) \cdot \mathbf{E}(\mathbf{r}_k(t),t)) (-\hat{d}(t') \cdot \mathbf{E}(\mathbf{r}_k(t'),t')) | 0 \rangle \right] =\frac {1} {4} \left[ g^R_{\hat{\mathbf{d}}}(t,t') \mathcal{G}_{\hat{\mathbf{E}}}^R(\mathbf{r}_k(t),t;\mathbf{r}_k(t'),t')+  g^R_{\hat{\mathbf{d}}}(t,t') \mathcal{G}_{\hat{\mathbf{E}}}^H(\mathbf{r}_k(t),t;\mathbf{r}_k(t'),t') \right]$. As the phase is real, one can indifferently write
\begin{equation}
\phi_k =  \frac {{i}} {2 \hbar^2} \int_{-T/2}^{T/2} dt \int_{-T/2}^{t} dt'  \left( \langle 0 | \hat{V}(\mathbf{r}_k(t),t) \hat{V}(\mathbf{r}_k(t'),t') | 0 \rangle - h.c. \right)
\end{equation}
Following the derivation in the Letter, the motional phase correction is obtained by performing a perturbative expansion in the retarded position $\mathbf{r}_k(t')=\mathbf{r}_k(t) - \tau \mathbf{\dot{r}}_k(t)$. This motional phase correction corresponds to the QVSP (by symmetry, for the considered trajectories, these motional corrections would vanish for a nano-particle at rest)
\begin{equation}
\phi_{k}^{\Omega} =  \frac {i} {{2} \hbar^2} \int_{-T/2}^{T/2} dt \: \dot{\mathbf{r}}_k(t)\cdot  \int_{0}^{t+T/2} d \tau  \: \tau \: \left( \langle 0 | \hat{V}(\mathbf{r}_k(t),t) \nabla_{\mathbf{r}} \hat{V}(\mathbf{r}_k(t),t-\tau) | 0 \rangle - h.c. \right) 
\end{equation}
where we have done the variable change $\tau=t-t'$. We then take the long-time limit $T \gg {\rm min}(1/\omega_n)$, use a closure relation, and switch back to the Schr\"odinger picture with Eq.~(\ref{eq:interaction_picture}). The QVSP can thus be written as
\begin{equation}
\phi_{k}^{\Omega}  =  \frac {i} {{2}  \hbar^2} \int_{\mathcal{P}_k} d \mathbf{r}\cdot \left(  \sum_{n \neq 0} \langle 0 | \hat{V}(\mathbf{r}) | n \rangle \langle n |  \nabla_{\mathbf{r}} \hat{V}(\mathbf{r}) | 0 \rangle  \int_0^{+\infty} d \tau  \: \tau \:  e^{-i (\omega_n-\omega_0) \tau}  - h.c. \right)
\end{equation}
We have used $d \mathbf{r} = dt \dot{\mathbf{r}}_k(t)$ to obtain a geometric integral along the path $\mathcal{P}_k$. The sum has been restricted to $n \neq 0$ without loss of generality as $ \langle 0 | \hat{V}(\mathbf{r})| 0 \rangle =0$. The integral over $\tau$ can be formally expressed as
 $$ \int_0^{+ \infty} d \tau  \: \tau \:  e^{-i (\omega_n-\omega_0) \tau}= -i \mathcal{P} \frac {1} {(\omega_n - \omega_0)} \left( i \mathcal{P} \frac {1} { (\omega_n-\omega_0)} + \pi \delta( \omega_n- \omega_0) \right) \, ,$$
where $\mathcal{P}$ is the principal part. The contribution of the Dirac distribution can be safely ignored as all the quantum states  ${|n \rangle, n>0}$ have an energy strictly above that initial state, with an energy difference corresponding at least to the two-level frequency. Finally, the QVSP can be written as:
\begin{equation}
\label{eq:motional_correction_final}
\phi_{k}^{\Omega} =  \frac {i} {{2} \hbar^2} \int_{\mathcal{P}_k} d \mathbf{r}\cdot   \sum_{n \neq 0} \frac {\left( \langle 0 | \hat{V}(\mathbf{r}) | n \rangle \langle n |  \nabla_{\mathbf{r}} \hat{V}(\mathbf{r}) | 0 \rangle    -  \langle 0 |  \nabla_{\mathbf{r}} \hat{V}(\mathbf{r}) | n \rangle \langle n |  \hat{V}(\mathbf{r}) | 0 \rangle  \right)
} {(\omega_n-\omega_0)^2}   \end{equation}

\subsection{Expression of the Berry phase of the full quantum system ``atom+field''}

 The full quantum system (internal atomic degrees of freedom+EM field) is steered as the atomic position evolves along the path $\mathcal{P}_k$, and acquires a geometric Berry phase resulting from the continuous modification of the ground-state $|\psi_0(\mathbf{r}) \rangle$:
\begin{equation}
\label{eq:Berryphase}
\phi_k^{\rm Berry} = i \int_{\mathcal{P}_k} d \mathbf{r}\cdot \langle \psi_0(\mathbf{r}) | \nabla_{\mathbf{r}} \psi_0(\mathbf{r}) \rangle 
\end{equation}
The atomic motion does not produce by itself no excitation of the internal atomic level for the considered non-relativistic atomic velocities: thus, the evolution is completely adiabatic. For the considered closed atom interferometers where the two trajectories ($\mathcal{P}_1,\mathcal{P}_2$) split at the initial time and recombine at the final time, the geometric phase contribution to the phase difference corresponds to the integration of the parameter $\mathbf{r}$ over a cycle.

We derive below this Berry phase to second-order in the interaction potential. Consistently with our calculation of the dispersive phase, we use standard-perturbation theory in order to express the instantaneous, position-dependent eigenstates in the interaction-free basis:
\begin{eqnarray}
| \psi_0( \mathbf{r} ) \rangle  & = & | 0 \rangle + \sum_{n \neq 0} \frac {\langle n | \hat{V}(\mathbf{r}) | 0 \rangle} {\hbar(\omega_n-\omega_0)} | n \rangle   - \frac 1 2 | 0 \rangle  \sum_{n \neq 0}\frac {|\langle n | \hat{V}(\mathbf{r}) | 0 \rangle|^2} {\hbar^2 (\omega_n-\omega_0)^2}
 \nonumber \\
& & + \sum_{n \neq 0, n' \neq n} 
| n \rangle \frac {\langle n | \hat{V}(\mathbf{r}) | n'\rangle \langle n' | \hat{V}(\mathbf{r}) | 0 \rangle} {\hbar^2 (\omega_0-\omega_n) (\omega_n-\omega_{n'})} -  \sum_{n \neq 0} | n \rangle   \frac {\langle n | \hat{V}(\mathbf{r}) |0 \rangle \langle 0 | \hat{V}(\mathbf{r}) | 0 \rangle} {\hbar^2 (\omega_n-\omega_0)^2}  +O \left( \hat{V}(\mathbf{r})^3 \right)
\end{eqnarray}
For the considered dipolar interaction the terms of the second line on the right-hand side actually vanish, as the dipole operator elements are null on the diagonal (at least two of the three system eigenstates $\{ | n \rangle, | n' \rangle, | 0 \rangle = | g \rangle_A \otimes | 0 \rangle_F \}$) involve the same atomic state).

One obtains:
\begin{equation}
| \nabla_{\mathbf{r}} \psi_0( \mathbf{r} ) \rangle   = \sum_{n \neq 0} \frac {\langle n | \nabla_{\mathbf{r}} \hat{V}(\mathbf{r}) | 0 \rangle} {\hbar(\omega_n-\omega_0)} | n \rangle   - \frac 1 2 | 0 \rangle  \sum_{n \neq 0}\frac {\langle 0 | \nabla_{\mathbf{r}}  \hat{V}(\mathbf{r}) | n \rangle \langle n | \hat{V}(\mathbf{r}) | 0 \rangle +\langle 0 |  \hat{V}(\mathbf{r}) | n \rangle \langle n | \nabla_{\mathbf{r}} \hat{V}(\mathbf{r}) | 0 \rangle } {(\omega_n-\omega_0)^2} +O \left( \hat{V}(\mathbf{r})^3 \right)
\end{equation}
yielding the Berry connection
\begin{equation}
i  \langle  \psi_0( \mathbf{r} ) \ | \nabla_{\mathbf{r}} \psi_0( \mathbf{r} ) \rangle = \frac {i} {2 \hbar^2} \sum_{n \neq 0} \left( \frac {\langle 0 | \hat{V}(\mathbf{r}) | n \rangle \langle n | \nabla_{\mathbf{r}}   \hat{V}(\mathbf{r}) | 0 \rangle - \langle 0 | \nabla_{\mathbf{r}}    \hat{V}(\mathbf{r}) | n \rangle \langle n |  \hat{V}(\mathbf{r}) | 0 \rangle } {(\omega_n-\omega_0)^2}\right) +O \left( \hat{V}(\mathbf{r})^3 \right)
\end{equation}
This shows that the QVSP (\ref{eq:motional_correction_final}) corresponds to the Berry phase~(\ref{eq:Berryphase}), i.e. $\phi_{k}^{\Omega} =\phi_k^{\rm Berry}$.

    \section{Nonlocal Quantum Vacuum Sagnac phase}

\subsection{General expression of the non-local QVSP.}


We obtain in this Section the non-local QVSP, which depends simultaneously on two paths $\alpha,\beta,$ and can be written as
 \begin{equation}
\phi_{\alpha \beta}^{\Omega}  =  \! \frac {- \omega_0 \alpha_0^{\rm A}  } {2} \int_{-T/2}^{T/2} \! \! \! \! \! \! dt  \:    \mathbf{v}_{\beta}(t) \cdot   \: \:   \nabla_{\mathbf{r}'}   \left. \frac {d} {d \omega} \mbox{Im} \left[  \delta \mathcal{G}^{R,S}_{\hat{\mathbf{E}}}(\mathbf{r},\mathbf{r}';\omega) \right]  \right|_{\mathbf{r}=\mathbf{r}_{\alpha}(t), \mathbf{r}'=\mathbf{r}_{\beta}(t),\omega=\omega_0}  
\label{eq:DPSagnac}
\end{equation}
for $\alpha  \neq \beta$. We neglect retardation effects, so that the non-local QVSP reads:
 \begin{eqnarray}
 \label{eq:SagnacDPgeral}
\phi^{\rm \Omega}_{\alpha \beta}  =  
 \frac {\alpha_0^A  \tilde{\alpha}''_R(\omega_0)} {2 (4 \pi \epsilon_0)^2} \int_{-T/2}^{T/2}  dt  \frac {\boldsymbol{\xi}_{\alpha \beta}(t)   \cdot \mathbf{\Omega}} {r_{\alpha}(t)^3 r_{\beta}(t)^4}    
 \end{eqnarray}  
with $\mathbf{\xi}_{\alpha \beta \: m}(t)= r_{\beta}(t) \sum_{i,j,k,l}   \epsilon_{klm}   v_{\beta i}(t)     T^0_{jk}(\mathbf{r}_\alpha(t))   \partial_{i}  T^0_{lj}(\mathbf{r}_\beta(t)).$ 
We omit the explicit time dependence in the positions and velocities in order to simplify notations. From Eq.\eqref{eq:vdWpropagator}, the quantity $ \boldsymbol{\xi}_{\alpha \beta}(t)$ can be expressed as
      \begin{equation}
 \xi_{\alpha \beta \: m}(t)=  3  \sum_{i,j,k,l}   \epsilon_{klm}       v_{\beta i}  \left( 3 \hat{r}_{\alpha j} \hat{r}_{\alpha k} - \delta_{jk} \right)   \left( \delta_{ij} \hat{r}_{\beta l} +  \delta_{il} \hat{r}_{\beta j}- 2 \hat{r}_{\beta i} \hat{r}_{\beta j} \hat{r}_{\beta l}  \right) \nonumber
 \end{equation}
  which can be put, after summation, in the simpler geometric form
  \begin{equation}
 \boldsymbol{\xi}_{\alpha \beta}(t)= \left[ 1- 2 (\hat{\mathbf{r}}_{\alpha} \cdot \hat{\mathbf{r}}_{\beta})   \right]  (\hat{\mathbf{r}}_{\alpha} \cdot \mathbf{v}_{\beta})  \hat{\mathbf{r}}_{\alpha} \times \hat{\mathbf{r}}_{\beta} + (\hat{\mathbf{r}}_{\alpha} \cdot \hat{\mathbf{r}}_{\beta}) \hat{\mathbf{r}}_{\alpha} \times \mathbf{v}_{\beta} 
 \label{eq:SagnacDPxl}
 \end{equation}
Eqs.~(\ref{eq:SagnacDPgeral},\ref{eq:SagnacDPxl}) constitute the general geometric form of the non-local QVSP.

  \subsection{Non-local QVSP for parallel linear trajectories}
 
 Let us derive the non-local QVSP in a particular geometry of two linear paths $\alpha,\beta$ equidistant to the spinning particle. We thus take  $\mathbf{r}_{1,2}(y) = \pm x_1 \hat{\mathbf{x}}+ y \hat{\mathbf{y}}$ with $x_1>0$ and $\mathbf{v}_{1,2}=v \mathbf{y}.$ One has $r_{1,2}(y)=r(y)=(x_1^2+y^2)^{1/2}$, $\hat{\mathbf{r}}_{\alpha} \cdot \hat{\mathbf{r}}_{\beta}=(y^2-x_1^2)/r(y)^2, $ $\hat{\mathbf{r}}_{\alpha} \cdot \hat{\mathbf{v}}_{\beta}=v y/r(y), $  $\hat{\mathbf{r}}_{\alpha} \times \mathbf{v}_{\beta}= (-1)^{\alpha+1} x_1 v/r(y) \mathbf{z}$, $\hat{\mathbf{r}}_{\alpha} \times \hat{\mathbf{r}}_{\beta}=(-1)^{\alpha+1} 2 x_1 y /r(y)^2 \hat{\mathbf{z}.}$
 Taking the limit $v T/2=L/2 \gg x_1,$ one obtains
  \begin{eqnarray}
\phi^{\rm \Omega}_{\alpha \beta} &   =   &
 \frac {9 (-1)^{\alpha+1}   \alpha_0^A  \tilde{\alpha}''_R(\omega_0) \Omega} {2 (4 \pi \epsilon_0)^2} \int_{-\infty}^{+\infty}  \frac {d y} {(x_1^2+y^2)^{7/2}} \left[ \frac {(3 x_1^2 -y^2) 2 x_1 y^2} {(x_1^2+y^2)^{5/2}} + \frac {(y^2-x_1^2)x_1} {(x_1^2+y^2)^{3/2}} \right] \nonumber \\
 &  =   &  - \frac {27 \pi (-1)^{\alpha+1}   \alpha_0^A  \tilde{\alpha}''_R(\omega_0) \Omega} {64 (4 \pi \epsilon_0)^2 x_1^6} 
 \end{eqnarray}

\section{Quantum Vacuum Sagnac phase for finite atomic wave packets}
\label{sec:averageQVSP}

We estimate the QVSP for finite atomic wave packets propagating at a constant velocity nearby the spinning nano-particle. We take for the atomic wave function $\psi(x,y,z,t)=   \psi_{\perp}(x,z) \psi_{/ \! /} (y,t)  $ with $\psi_{\perp}(x,z)= \frac {1} {\pi w^2} e^{-\frac {(x-a)^2} {2 w^2}}   e^{ -\frac {z^2} {2 w^2}}.$ and $\psi_{/ \! /} (y,t) =\frac {1} {\pi^{1/2} w_y}  e^{-\frac {(y-vt)^2} {2 w_y^2}+ \frac {i} {\hbar}  m v y},$ where $m$ is the Sodium mass. This corresponds to a CM atomic motion at a constant velocity $v$ along a line $\{ x=a, y=vt, z=0 \}$ passing at the edge of the spinning nano-particle, and with a constant transverse profile. This approximation is legitimate as the time of flight is in practice too small to yield a noticeable change in the wave-packet shape. Indeed for the considered transition wave-length $\lambda_0$ and velocity $v$, the time-of flight is approximately $T  \sim \lambda_0 / v \sim 10^{-10} \rm {\: s.}$ Even for the smallest considered transverse width $w \sim 8 {\rm nm}$, the transverse velocity dispersion $\Delta v_{\perp} = \hbar / m w \sim 0.4 \: {\rm m/s}$ results in a negligible increase of the wave packet size during $T$. The experimentally accessible phase is obtained through a WKB approach and an averaging procedure as exposed in Refs.~\cite{Cronin04,Vigue11}. Here, one proceeds to an averaging over the bundle of linear trajectories $\{ \mathbf{r}(t)=x \hat{\mathbf{x}}+ v t \hat{\mathbf{y}}+ z \hat{\mathbf{z}} \}$ with a weight given by the transverse wave-function profile. The resulting phase is given by
\begin{equation}
\overline{\phi}(\Omega,v)={\rm Arctan} \left[ \frac {\overline{\sin( \phi(\Omega,x,z,v) )}}  {\overline{\cos( \phi(\Omega,x,z,v) )}}  \right] \, .
\label{eq:Cornuintegration}
\end{equation}
Here $ \phi(\Omega,x,z,v)$ is the sum of the quasistatic vdW and Sagnac phases, namely $\phi(\Omega,x,z,v) = \phi^{\rm vdW}(x,z,v)+ \phi^{\Omega}(x,z,v).$ The averaging is defined as $\overline{f(x,z,...)} = \int dx dz f(x,z,...) | \psi_{\perp}(x,z) |^2$. The wave-packet dispersion in the longitudinal coordinate $y$ does not affect the dispersive phase. The QVSP acquired along a line $\{ x'=x, - L/2 \leq y' \leq L/2 ,z'=z \}$ reads $\phi^{\Omega}(x,z,v)=45 \pi \ell^6_{\Omega}\, x/ (32 r_{\perp}^7) $ with $r_{\perp}=(x^2+z^2)^{1/2}$ and $\ell_{\Omega}= [\omega_0 \alpha_0 \alpha_R''(\omega_0)  \Omega / (4 \pi \epsilon_0)^2]^{1/6}.$ The quasistatic vdW phase depends on the velocity $v$ and corresponds to ($\overline{\alpha}_I(\omega_0) \simeq \tilde{\alpha}_R(\omega_0)$ within a $1\%$ approximation) $$\phi^{\rm vdW}(x,z,v) = \frac {9 \pi \alpha_0 \omega_0 (\tilde{\alpha}_R(\omega_0)+\overline{\alpha}_I(\omega_0))} {8 (4 \pi \epsilon_0)^2  v r_{\perp})^{5} } \simeq  \frac {9 \pi \alpha_0 \omega_0 \tilde{\alpha}_R(\omega_0)} {4 (4 \pi \epsilon_0)^2 v r_{\perp}^{5}} $$

\end{widetext}


\begin{thebibliography}{12}


\bibitem{Sagnac13} G.~Sagnac, C.R. Acad. Sci. Paris~\textbf{141}, 1220 (1913); G.~Sagnac, C.R. Acad. Sci. Paris~ \textbf{157}, 708 (1913).

\bibitem{Borde89} Ch. J. Bord\'e,  Phys. Lett. A,~\textbf{140}, 10, (1989).


\bibitem{Gustavson96} T. L. Gustavson, P. Bouyer, and M. A. Kasevich,
Phys. Rev. Lett.~\textbf{78}, 2046 (1996).

\bibitem{Pritchard97} A. Lenef, T. D. Hammond, E. T. Smith, M. S. Chapman, R. A. Rubenstein, and D. E. Pritchard,
Phys. Rev. Lett.~\textbf{78}, 760 (1997).

\bibitem{Canuel06} B. Canuel, F. Leduc, D. Holleville, A. Gauguet, J. Fils, A. Virdis, A. Clairon, N. Dimarcq, Ch. J. Bord\'e, A. Landragin, and P. Bouyer,
Phys. Rev. Lett.~\textbf{97}, 010402 (2006).

\bibitem{RMPCronin09} A. D. Cronin, J. Schmiedmayer, and D. E. Pritchard,
Rev. Mod. Phys.~\textbf{81}, 1051 (2009).

\bibitem{Sackett20} E. R. Moan, R. A. Horne, T. Arpornthip, Z. Luo, A. J. Fallon, S. J. Berl, and C. A. Sackett,
Phys. Rev. Lett.~\textbf{124}, 120403 (2020).

\bibitem{Ryu2020} C. Ryu, E. C. Samson, and M. G. Boshier,
 Nat. Commun.~\textbf{11}, 3338 (2020).

\bibitem{Schubert2021} C. Schubert, S. Abend, M. Gersemann, M. Gebbe, D. Schlippert, P. Berg, and E. M. Rasel,
Sci. Rep.~\textbf{11}, 16121  (2021).

\bibitem{Geiger2020}  R. Geiger,  A. Landragin,  S. Merlet, and  F. Pereira Dos Santos,
AVS Quantum Sci. \textbf{2}, 024702 (2020).

\bibitem{Ahn2020} 
J. Ahn, Z. Xu, J. Bang, P. Ju, X. Gao, and T. Li,
 Nat. Nanotechnol. {\bf 15}, 89 (2020). 

\bibitem{Dalvit2011} D. A. R. Dalvit, P. A. Maia Neto and F. D. Mazzitelli, Lect. Notes Phys. \textbf{834}, 287 (2011).

\bibitem{Dodonov2020} V. Dodonov,
Physics {\bf 2}, 67 (2020).

 \bibitem{MaiaNeto96} P.  A. Maia Neto  and L.  A.  S.  Machado,
 Phys. Rev. A {\bf 54}, 3420 (1996).
 
\bibitem{Dodonov2021} L. Woods, M. Kr\"uger, and V.V.Dodonov,  Appl. Sci. {\bf 11}, 293 (2021).

\bibitem{Dalvit2021} D. A. R. Dalvit and W. Kort-Kamp, Universe {\bf 7}, 189 (2021). 

\bibitem{Lambrecht96} A. Lambrecht,  M.-T. Jaekel,  and S. Reynaud, 
Phys. Rev. Lett. {\bf 77}, 615 (1996). 

\bibitem{Dodonov96} V. V. Dodonov and A. B. Klimov, Phys. Rev. A {\bf 53}, 2664 (1996).

\bibitem{Schutzhold98}  R. Sch\"utzhold,  G. Plunien  and  G. Soff, Phys. Rev. A {\bf 57}, 2311
(1998). 

\bibitem{Crocce2001} M. Crocce, D. A. R. Dalvit, and F. D. Mazzitelli, 
Phys. Rev. A {\bf 64}, 013808 (2001). 

\bibitem{Kumiya2016} T. Kumiya,  A. S. Akentyev, Y. Mori, J. Ichimura, and A. Morinaga, Phys. Rev. A {\bf 93}, 023637 (2016).

\bibitem{Reimann2018} R. Reimann, M. Doderer, E. Hebestreit, R. Diehl, M. Frimmer, D. Windey, F. Tebbenjohanns, and L. Novotny,
 Phys. Rev. Lett. {\bf 121}, 033602 (2018). 

\bibitem{Ahn2018} J. Ahn, Z. Xu, J. Bang, Y.-H. Deng, T. M. Hoang, Q. Han, R.-M. Ma, and T. Li,
Phys. Rev. Lett. {\bf 121}, 033603 (2018). 


\bibitem{Barton96} G. Barton, 
Ann. Phys.  (N.Y.) {\bf 245}, 361 (1996). 

\bibitem{Manjavacas2010} A. Manjavacas and F. J. Garc\'{i}a de Abajo. Phys. Rev. A~\textbf{82}, 063827 (2010).

\bibitem{ManjavacasPRL2010} A. Manjavacas and F. J. Garc\'{\i}a de Abajo, 
Phys. Rev. Lett.~\textbf{105}, 113601 (2010). 

\bibitem{Nonlocal1} F. Impens, R. O. Behunin, C. C. Ttira and P. A. Maia Neto, Eur. Phys. Lett.~\textbf{101}, 60006 (2013).

\bibitem{Nonlocal2} F. Impens, C. C. Ttira, and P. A Maia Neto, J. Phys. B: At. Mol. Opt. Phys.~\textbf{46}, 245503 (2013).

\bibitem{Nonlocal3} F. Impens,  C. C. Ttira, R. O. Behunin, and P. A. Maia Neto,
Phys. Rev. A~\textbf{89}, 022516 (2014).

\bibitem{Souza2018}  R. M. Souza, F. Impens,  and P. A. Maia Neto,
Phys. Rev. A~\textbf{97}, 032514 (2018).

\bibitem{Lo2018} L. Z. Lo and C. K. Law, Phys. Rev. A {\bf 98}, 063807 (2018). 

\bibitem{Farias2019} M. B. Far\'{\i}as, C. D. Fosco, F. C. Lombardo, and F. D. Mazzitelli, Phys. Rev. D {\bf 100}, 036013 (2019). 



\bibitem{Dolan2020} B. P. Dolan, A. Hunter-McCabe, and J. Twamley, New J. Phys. {\bf  22}, 033026 (2020). 

\bibitem{Agusti2021} A. Agust\'{\i}, L. Garc\'{\i}a-\'Alvarez, E. Solano, and C. Sab\'{\i}n,
Phys. Rev. A {\bf  103}, 062201 (2021). 

\bibitem{Scheel2012} S. Scheel and S. Y. Buhmann, Phys. Rev. A~\textbf{85}, 030101(R) (2012).

\bibitem{Souza2016} 
R. Melo Souza, F. Impens, and P. A. Maia Neto, 
Phys. Rev. A {\bf 94}, 062114 (2016). 


\bibitem{Scheel09} S. Scheel, S. Y. Buhmann, Phys. Rev. A~\textbf{80}, 042902 (2009).

\bibitem{Pieplow2013} G. Pieplow and C. Henkel, New J. Phys. {\bf 15}, 023027  (2013).

\bibitem{Donaire2016}  M. Donaire and A. Lambrecht, Phys.Rev.A~\textbf{93}, 022701 (2016).

\bibitem{Reiche2020} D. Reiche, F. Intravaia, J.-T. Hsiang, K. Busch, and B. L. Hu, Phys. Rev. A {\bf 102}, 050203(R) (2020). 

\bibitem{Farias2020} M. B. Farias, F. C. Lombardo, A. Soba, P. I. Villar and R. S. Decca, npj Quantum Inf~\textbf{6}, 25 (2020).

\bibitem{Fosco2021} C. D. Fosco, F. C. Lombardo, F. D. Mazzitelli, Universe {\bf 7}, 158 (2021)
 
 \bibitem{Lombardo21} F. C. Lombardo, R. S. Decca, L. Viotti, P. I. Villar, Adv. Quant. Tech~\textbf{4}, 2000155 (2021).

\bibitem{Aharonov1959} Y. Aharonov, and D. Bohm,  
Phys. Rev. {\bf 115} 485 (1959). 

\bibitem{ReviewSagnacBerryArxiv21} Ismael L. Paiva, Rain Lenny, Eliahu Cohen, e-print arXiv:2110.05824 (2021).

\bibitem{Dalibard11} J. Dalibard, F. Gerbier, G. Juzeli\"unas, and P. \"Ohberg,
Rev. Mod. Phys.~\textbf{83}, 1523 (2011).

\bibitem{Cornell04} V. Schweikhard, I. Coddington, P. Engels, V. P. Mogendorff, and E. A. Cornell,
Phys. Rev. Lett.~\textbf{92}, 040404 (2004).

\bibitem{Miniatura92} Ch. Miniatura, J. Robert, O. Gorceix, V. Lorent, S. Le Boiteux, J. Reinhardt, and J. Baudon,
Phys. Rev. Lett.~\textbf{69}, 261 (1992). 

\bibitem{Gauguet13} J. Gillot, S. Lepoutre, A. Gauguet, M. B\"uchner and J. Vigu\'e,  Phys. Rev. Lett.~\textbf{111}, 030401 (2013).

 \bibitem{PlasmonMetallic2012} J. A. Scholl, A. L. Koh, and J. A. Dionne, Nature~\textbf{483}, 421 (2012).

\bibitem{HeitlerBook} W. Heitler, \textit{The Quantum Theory of Radiation},  Chapt. II, (Dover, New York, 1954).

\bibitem{Lombardo06} F. C. Lombardo, P. I. Villar,
Phys. Rev. A~\textbf{74}, 042311 (2006).

\bibitem{Supplementary} See supplementary material.

\bibitem{Berry1} M. V. Berry, Proc. R. Soc. Lond. A~\textbf{392}, 45 (1984).

\bibitem{Berry2} A. Shapere and F. Wilczek,~\textit{Geometric Phases in Physics}, (World Scientific, Singapore,1989).

\bibitem{Cronin04} A. D. Cronin and J. D. Perreault, Phys. Rev. A~\textbf{70}, 043607 (2004).

\bibitem{Vigue09} S. Lepoutre, H. Jelassi, V. P. A. Lonij, G. Tr\'enec, M. B\"uchner, A. D. Cronin and J. Vigu\'e, Eur. Pys. Lett.~\textbf{88} 20002 (2009).

\bibitem{Vigue11} S. Lepoutre, V. P. A. Lonij, H. Jelassi, G. Tr\'enec, M. B\"uchner, A.D. Cronin, and J. Vigu\'e,  Eur. Phys. J. D~\textbf{62}, 309 (2011).

\bibitem{Mazzitelli03} Francisco D. Mazzitelli, Juan Pablo Paz, and Alejandro Villanueva,
Phys. Rev. A~\textbf{68}, 062106 (2003).

\bibitem{Cronin10} W. F. Holmgren, M. C. Revelle, V. P. A. Lonij, and A. D. Cronin, Phys. Rev. A~\textbf{81}, 053607 (2010).

\bibitem{Derkachova2016} A. Derkachova, K. Kolwas, and  I. Demchenko, Plasmonics {\bf 11}, 941 (2016). 

\bibitem{Karimi2019} S. Karimi, A. Moshaii, S. Abbasian, and M. Nikkhah, Plasmonics {\bf 14}, 851 (2019). 

\bibitem{Blaber09} M. G. Blaber,  M. D. Arnold, and M. J. Ford, J. Phys. Chem. C~\textbf{113}, 3041 (2009).

\bibitem{Prentiss92} G. Timp, R. E. Behringer, D. M. Tennant, J. E. Cunningham, M. Prentiss, and K. K. Berggren,
Phys. Rev. Lett.~\textbf{69}, 1636 (1992). 

\bibitem{Martin21a} R. Richberg, S.S. Szigeti, A.M. Martin, Phys. Rev. A~\textbf{103}, 063304 (2021).

\bibitem{Martin21b} R. Richberg, A. M. Martin, Phys. Rev. A~\textbf{104}, 033314 (2021). 

\bibitem{Salvi2018} L. Salvi, N. Poli, V. Vuleti\'c, and G. M. Tino, Phys. Rev. Lett. {\bf 120}, 033601 (2018).


\bibitem{ThiruBook} D. P. Craig, T. Thirunamachandran, \textit{Molecular Quantum Electrodynamics}, (Academic Press, London, 1984).


\end{thebibliography}
\end{document}